\documentclass[%
 reprint,
preprint,
 amsmath,amssymb,
 aps,showkeys,pre
]{revtex4-2}

\usepackage{graphicx}
\usepackage{dcolumn}
\usepackage{subfigure}
\usepackage{amsmath}
\usepackage{amssymb}
\usepackage{bm}

\begin{document}

\title{Generalized $XY$ model with competing antiferromagnetic and antinematic interactions}
\author{Milan \v{Z}ukovi\v{c}}
 \email{milan.zukovic@upjs.sk}
 \affiliation{Institute of Physics, Faculty of Science, P. J. \v{S}af\'arik University, Park Angelinum 9, 041 54 Ko\v{s}ice, Slovakia}
\date{\today}

\begin{abstract}

We study effects of $q$-order antinematic (AN$_q$) interactions on the critical behavior of the antiferromagnetic (AF) $XY$ model on a square lattice. It is found that the evolution of the phase diagram topology of such AF-AN$_q$ models with the parameter $q$ does not follow the same line as for the corresponding FM-N$_q$ models with the ferromagnetic (FM) and $q$-order nematic (N$_q$) interactions. Their phase diagrams are similar only for odd values of the parameter $q$. In such cases the respective phases reported in the FM-N$_q$ models are observed in the AF-AN$_q$ models on each of the two AF-coupled sublattices and the corresponding phase transitions are concluded to be of the same kind. On the other hand, for even values of $q$ the phase diagrams of the AF-AN$_q$ models are different from the FM-N$_q$ models and their topology does not change with $q$. Besides the pure AF and AN$_q$ phases, observed at higher temperatures in the regions of the dominant respective couplings, at low temperatures there is a new canted (C)AF phase, which results from the competition between the AF and AN$_q$ ordering tendencies and has no counterpart in the FM-N$_q$ model. The phase transitions to the CAF phase from both AF and AN$_q$ phases appear to be of the BKT nature.

\end{abstract}





\maketitle

\section{Introduction}

The standard two-dimensional $XY$ model is very well understood but its various generalizations are still of great interest due to their rich and unexpected critical behavior as well as their potential experimental realizations. In particular, the model with a nematic term has attracted a lot of attention as a potential model for various experimental realizations, including liquid crystals~\cite{lee85,carp89,geng09}, superfluid A phase of $^3{\rm He}$~\cite{kors85}, or high-temperature cuprate superconductors~\cite{hlub08}. A central feature of such a model is the existence of a nematic quasi-long-range order (QLRO) phase, which is separated from the magnetic one at lower temperatures by the phase boundary belonging to the Ising universality class~\cite{lee85,kors85}. Later it has been found that replacing the second-order nematic term with a more general $q$-order (pseudo-nematic) term, can result in even more interesting critical behavior. Namely, for $q \geq 3$ new ordered phases can appear and the phase transitions between different phases can belong to various, such as the Berezinskii-Kosterlitz-Thouless (BKT), the Ising or the three-state Potts universality classes~\cite{pode11,cano14,cano16}. 

Further generalizations, that were motivated by orientational transitions in liquid crystals, demonstrated possibility of the change of the BKT to the first-order transition. Such a scenario has been suggested and rigorously proved in the models with the $k$-th order Legendre polynomials of the dipole term for sufficiently large value of $k$~\cite{ente02,ente05}, the non-linear models with the potential shape controlled by the parameter $p^2$ for large $p$~\cite{doma84,himb84,blot02,sinh10a,sinh10b}, or the $XY$ models involving $q$ higher-order terms with exponentially vanishing strength for sufficiently large $q$ (including an infinite number)~\cite{zuko17}. Multiple phase transitions have been found in the models with a small number of purely low-order pseudo-nematic terms~\cite{zuko18a,zuko18b}. 

The corresponding models with the antiferromagnetic (AF) and $q$-order antinematic (AN$_q$) couplings have been mainly studied on a frustration-inducing non-bipartite triangular lattice. The model with $q=2$ has been demonstrated to produce the phase diagram, which besides the AF and AN$_2$ QLRO phases also includes a so-called chiral LRO phase with the phase boundary decoupled from the magnetic and nematic ones~\cite{park08}. More recent investigations of the AF-AN$_q$ models with $q>2$ on a triangular lattice led to the conclusion that such models can display a number of ordered and quasi-ordered phases as a result of geometrical frustration and/or competition between the AF and the AN$_q$ interactions~\cite{zuko16,lach20,lach21}. It is worth noting that such models have also been employed in some interdisciplinary applications, e.g. for modeling of DNA packing~\cite{gras08} or structural phases of cyanide polymers~\cite{cair16,clark16,zuko16}. 

Recent studies demonstrated that also the $XY$ model with mixed FM and AN$_2$ couplings on a (unfrustrated bipartite) square lattice is of theoretical interest as it shows a unique critical behavior~\cite{dian11,zuko19}. The resulting phase diagram is different from either the corresponding FM-AN$_2$ model on the triangular lattice or the FM-N$_2$ model on the same square lattice. Namely, it features a peculiar canted FM (CFM) phase that arises due to the competition between the FM and AN$_2$ couplings. In the present study, we systematically investigate effects of the AN$_q$ terms on the critical behavior of the AF $XY$ model on the square lattice for arbitrary value of the parameter $q$.

\section{Model and method}
We consider the generalized $XY$ models with the AF interaction $J_1$ and the AN$_q$ interaction $J_q$ on the square lattice with the Hamiltonian 
\begin{equation}
\label{Hamiltonian}
{\mathcal H}=-J_1\sum_{\langle i,j \rangle}\cos(\phi_{i,j})-J_q\sum_{\langle i,j \rangle}\cos(q\phi_{i,j}),
\end{equation}
where $\phi_{i,j}=\phi_{i}-\phi_{j}$ is an angle between nearest-neighbor spins, $J_1 \in (-1,0)$ and $J_q = -J_1-1 \in (-1,0)$, and $q$ is a positive integer. It is important to remark that in such models there is a competition between the two terms: while the first one favors the colinear antiparallel alignment of the neighboring spins, the second one enforces a phase difference of $2k\pi/q$, where $k \leq q$ is an integer.

To determine the ground states in the $J_q-q$ parameter space one needs to find the spin configurations that minimize the Hamiltonian in the phase space, which can be done by global optimization of the energy functional $\mathcal H$. To obtain temperature dependencies of various quantities of interest, we perform Monte Carlo (MC) simulation with Metropolis dynamics. We simulate lattices of the size $L\times L$, with the side length ranging between $L=24$-$120$ by applying periodic boundary conditions. For thermal averaging we typically consider $5 \times 10^5$ MC sweeps (MCS) after discarding the initial $10^5$ MCS necessary for bringing the system to the thermal equilibrium. The simulations proceed from high temperatures, corresponding to a paramagnetic phase, towards lower values. To shorten the thermalization period and to make sure that the system remains close to the equilibrium at all times the simulation at the next temperature is initiated using the final configuration obtained at the previous temperature separated from the current one by a sufficiently small step (typically $\Delta T=0.025$, measured in units of $J_1$ with the Boltzmann constant set to unity). 

To identify the nature (universality class) of a phase transition we need to perform a finite-size scaling (FSS) analysis. For that purpose it is useful to perform much longer runs at one temperature sufficiently close to the transition point for a wider range of the lattice sizes. Then by applying reweighting techniques~\cite{ferr88,ferr89} one can achieve a more accurate determination of the maxima of various quantities involved in the FSS analysis with the goal to estimate the corresponding critical exponents. For the FSS analysis we perform simulations using up to $5 \times 10^6$ MCS after discarding $10^6$ MCS for thermalization. The calculated mean values are accompanied with the statistical errors evaluated using the $\Gamma$-method~\cite{wolf04}.

We evaluate the following thermodynamic functions, where $\langle\cdots\rangle$ denotes thermal averaging. The specific heat per spin $c$ is obtained from the energy fluctuations as
\begin{equation}
c=\frac{\langle {\mathcal H}^{2} \rangle - \langle {\mathcal H} \rangle^{2}}{L^2T^{2}}.
\label{c}
\end{equation}
We define the sublattice order parameters $m_{k,l}$ as
\begin{equation}
m_{k,l}=\langle M_{k,l} \rangle/L^2=\left\langle\Big|\sum_{j}\exp(\mathrm{i}k\phi_j)\Big|\right\rangle/L^2,
\label{m}
\end{equation}
where $k=1,\hdots,q$ and $l=1,2$ denotes the two sublattices of the square lattice and the summation runs over the spins belonging to the sublattice $l$. We note that the values of $k=1$ ($k>1$) correspond to the magnetic (nematic) order parameters.
The corresponding sublattice susceptibilities $\chi_{k,l}$ are then defined as
\begin{equation}
\chi_{k,l} = \frac{\langle M_{k,l}^{2} \rangle - \langle M_{k,l} \rangle^{2}}{L^2T},\ k=1,\hdots,q,\ l=1,2.
\label{chi_q}
\end{equation}
The total order parameters and susceptibilities are then obtained as $m_k=m_{k,1}+m_{k,2}$ and $\chi_k=\chi_{k,1}+\chi_{k,2}$, respectively.

The order parameter for a BKT transition~\cite{berez71,kt73} is the helicity modulus $\Upsilon$ or spin wave stiffness~\cite{fish73,nels77,minn03,hubs13}. It reflects the response of the system upon a small overall twist of spins in a particular direction and in the thermodynamic limit it acquires the zero value in the disordered phase and finite values in the ordered phase. On the sublattice $l$ it is defined as
\begin{equation}
\label{helicity}
\begin{aligned}
\Upsilon_{l} = & \frac{1}{L^2}\left\langle \sum_{\langle i,j \rangle_x} J_1\cos (\phi_{i,j}) + q^2J_q\cos (q\phi_{i,j})\right\rangle \\ 
           & - \frac{\beta}{L^2}\left\langle\Big[\sum_{\langle i,j \rangle_x} J_1\sin (\phi_{i,j}) + qJ_q\sin (q\phi_{i,j})\Big]^2\right\rangle,
\end{aligned}
\end{equation}
where $\sum_{\langle i,j \rangle_x}$ is taken over the nearest neighbors on the sublattice $l$ along the twist direction. The total helicity modulus is then taken as $\Upsilon = \Upsilon_{1} + \Upsilon_{2}$.

We also evaluate a vortex density $\rho$, calculated directly from MC states. A vortex (antivortex) is a topological defect which corresponds to the spin angle change by $2\pi$ $(-2\pi)$ going around a closed contour enclosing the excitation core. The vortex density can be obtained by summation of the angles between adjacent four spins on each square plaquette for each equilibrium configuration. Then the value of $2\pi$ represents a vortex, $-2\pi$ antivortex and $0$ means that there is no topological defect. Finally, the vortex density $\rho$ is obtained as a normalized thermodynamic average of the absolute value of the vorticity (taking into consideration both vortices and antivortices) summed over the entire lattice and normalized by its volume.

\section{Results}
\subsection{Ground state}

\begin{figure*}[t!]
\centering
\subfigure{\includegraphics[scale=0.55,clip]{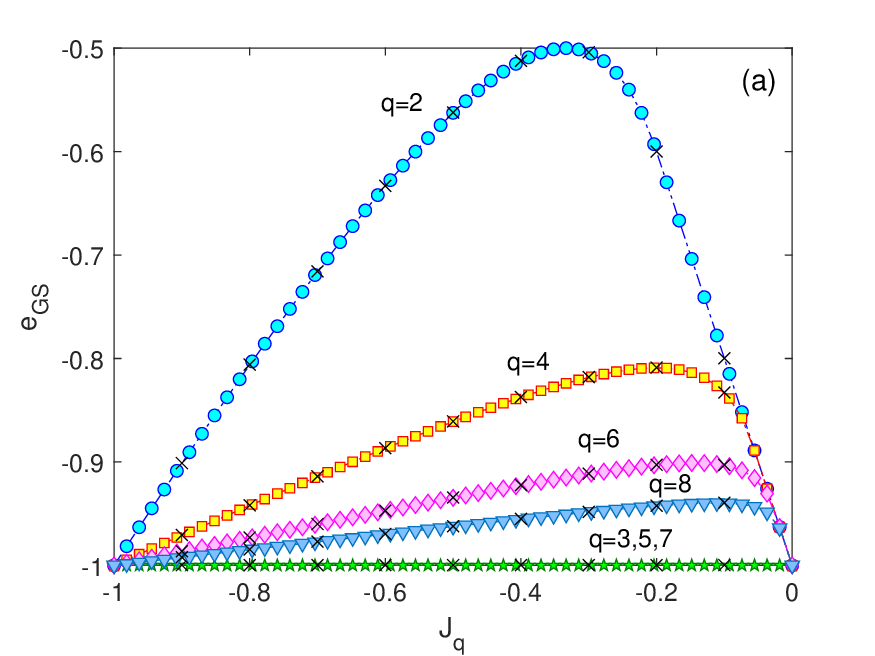}\label{fig:e_GS_q2-8}}
\subfigure{\includegraphics[scale=0.55,clip]{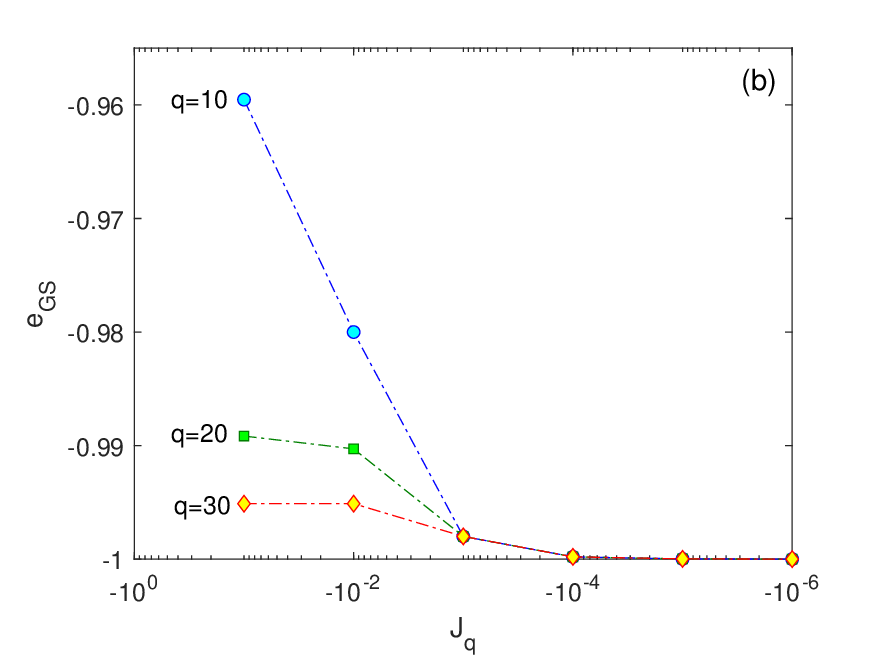}\label{fig:e_GS_det}}\\
\subfigure{\includegraphics[scale=0.55,clip]{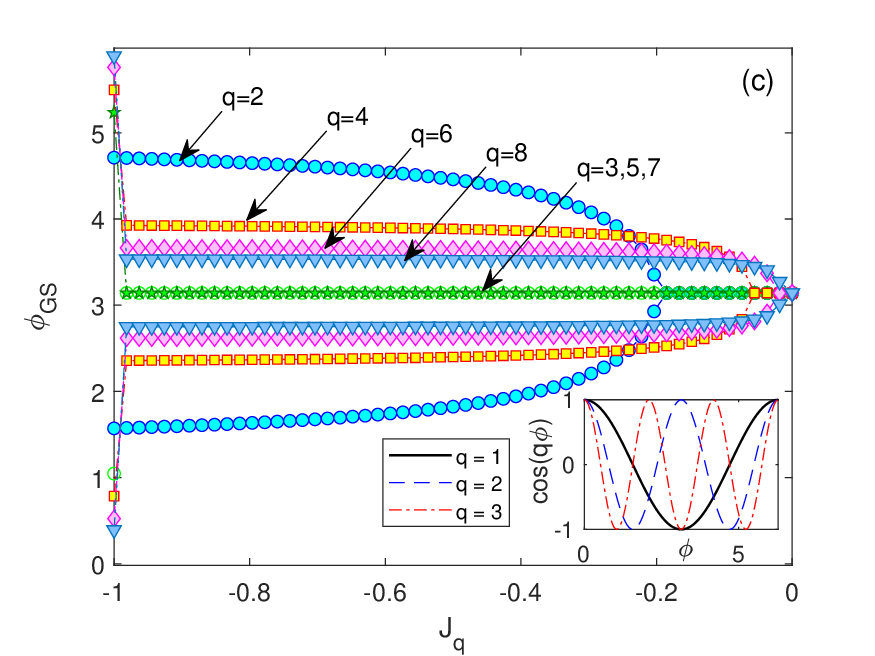}\label{fig:phi_GS_q2-8}}
\subfigure{\includegraphics[scale=0.55,clip]{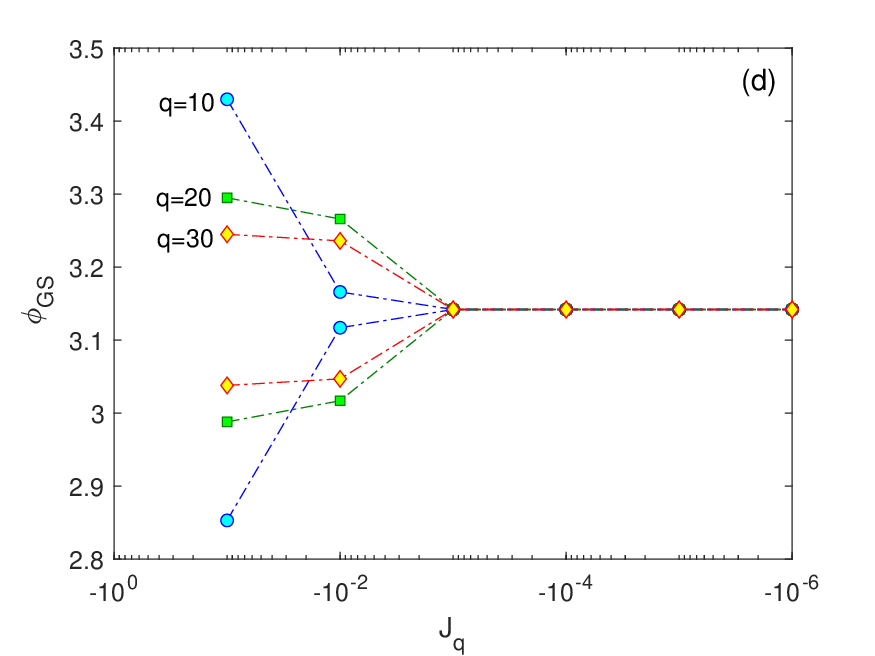}\label{fig:phi_GS_det}}
\caption{(Color online) (a,b) Ground-state energies per spin pair and (c,d) the corresponding spin angles, obtained from numerical optimization for different values of $q$. The black crosses in (a) show the energy per spin pair obtained from MC simulations at the lowest considered temperature. The panels (b) and (d) zoom in the behavior of the quantities for large $q$ in the limit of $J_q \to 0$.}\label{fig:gs}
\end{figure*}

Ground-state properties with the varying exchange interaction ratio are presented in Fig.~\ref{fig:gs}, for the increasing values of the parameter $q$. In particular, Fig.~\ref{fig:e_GS_q2-8} shows the variation of the ground-state energy per spin pair $e_{GS}$ with $J_q$, for $q=2,3,\hdots,8$. One can notice different behavior of the curves corresponding to even and odd values of $q$. All start and end at the minimum value $e_{GS}=-1$ but for even values there is an increase within the interval $J_q \in (-1,0)$ with the maximum of $e_{GS}=-0.5$ for $q=2$, which gradually decreases with the increasing $q$. The semi-log plot in Fig.~\ref{fig:e_GS_det} indicates that the enhanced energies persist up to large values of $q$ but not within the entire interval $J_q \in (-1,0)$. The energy seems to be minimized to $e_{GS}=-1$ before the limiting value of $J_q=0$, i.e., even in the presence of a sufficiently small $J_q \gtrsim -10^{-4}$. On the other hand, the ground-state energies for odd values of $q$ remain constant at $e_{GS}=-1$ within the entire interval $J_q \in (-1,0)$.

The bottom panels in Fig.~\ref{fig:gs} present the corresponding variations of the ground-state turn angles between neighboring spins $\phi_{GS}$ with $J_q$. The limiting cases of $\phi_{GS}=\pi$ at $J_q=0$ and $\phi_{GS}=\pm \pi/q$ at $J_q=-1$ correspond to the purely AF and AN$_q$ orderings, respectively, with no competition between the AF and AN$_q$ couplings and thus the minimum energy. Within $J_q \in (-1,J_q^*)$ for even values of $q$ there is competition between the two terms that leads to the non-universal ($J_q$-dependent) spin turn angles $\pi-\pi/q < \phi_{GS} < \pi+\pi/q$ that minimize the total energy at the values $e_{GS}>-1$. Hereafter, we will refer to this phase as canted antiferromagnetic (CAF). In the remaining part of the interval, i.e. within $J_q \in (J_q^*,0)$, there is still competition present but the $J_1$ term prevails, which leads to the perfect AF ordering, albeit at still elevated $e_{GS}$ (except the region close to $J_q = 0$, as shown in Fig.~\ref{fig:e_GS_det}). The threshold value $J_q^*$ increases with $q$ but for $q \to \infty$ does not vanish but instead it seems to converge to some small value of $J_q^* \approx -10^{-3}$ (see Fig.~\ref{fig:phi_GS_det}). 

The presence/absence of the competition for even and odd values of $q$ is demonstrated in the inset of Fig.~\ref{fig:phi_GS_q2-8}, which shows the energies of the AF ($q=1$) and AN$_q$ ($q>1$) terms as functions of the phase angle. While the minima for odd values of $q>1$ include the minimum of the $q=1$ phase at $\phi=\pi$ with $e_{GS}=-1$, those for even $q$, such as $q=2$, are shifted leading to the enhanced $e_{GS}>-1$ at $\phi \neq \pi$, corresponding to the CAF phase. Within the CAF phase we observed formation of small domains with AF (antiparallel) orientation of spins the size of which gradually increases with the increasing value of $J_q$. Eventually, for $J_q > J_q^*$ the canting angle $\phi_{GS}$ becomes $\pi$ and the domains merge to a single AF domain spanning the entire lattice. The spin domains observed in the CAF phase have zero-energy walls due to the inherent degeneracy caused by the competing interactions (see Ref.~\cite{zuko19} for details in the FM-AN$_2$ case).

\subsection{Finite temperatures}
\subsubsection{Phase diagrams}

\begin{figure*}[t!]
\centering
\subfigure{\includegraphics[scale=0.55,clip]{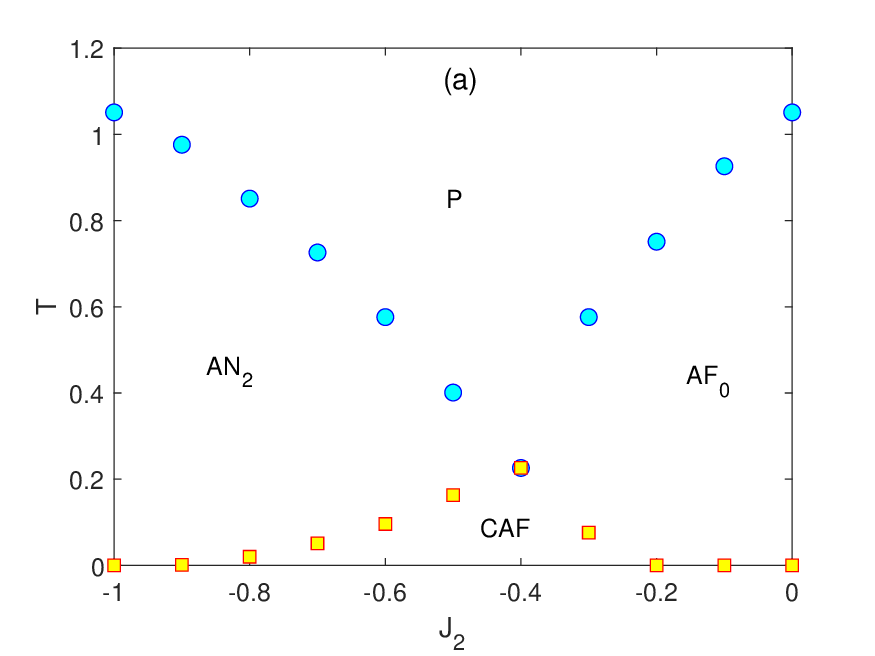}\label{fig:PD_q2}}
\subfigure{\includegraphics[scale=0.55,clip]{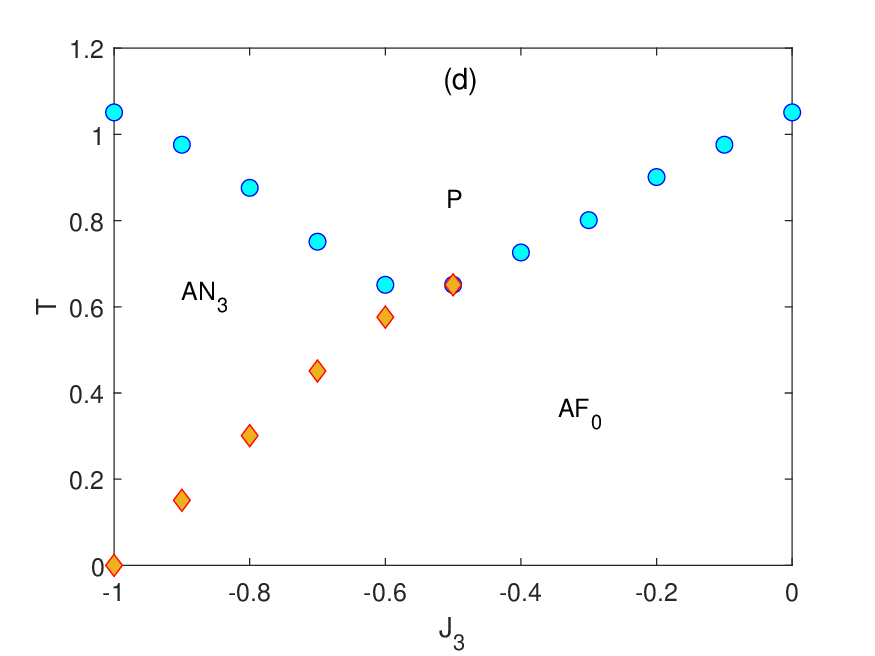}\label{fig:PD_q3}} \\
\subfigure{\includegraphics[scale=0.55,clip]{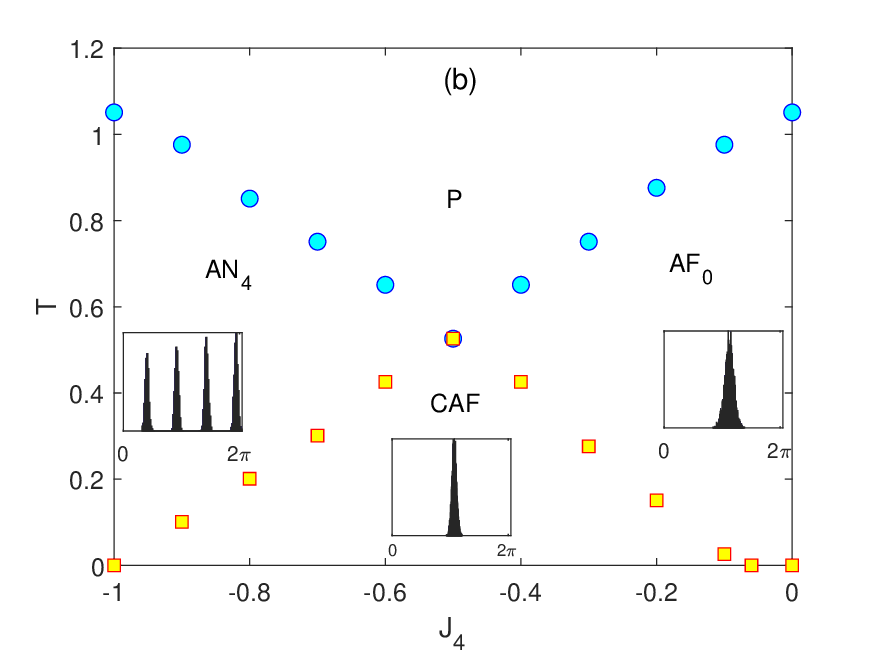}\label{fig:PD_q4}} 
\subfigure{\includegraphics[scale=0.55,clip]{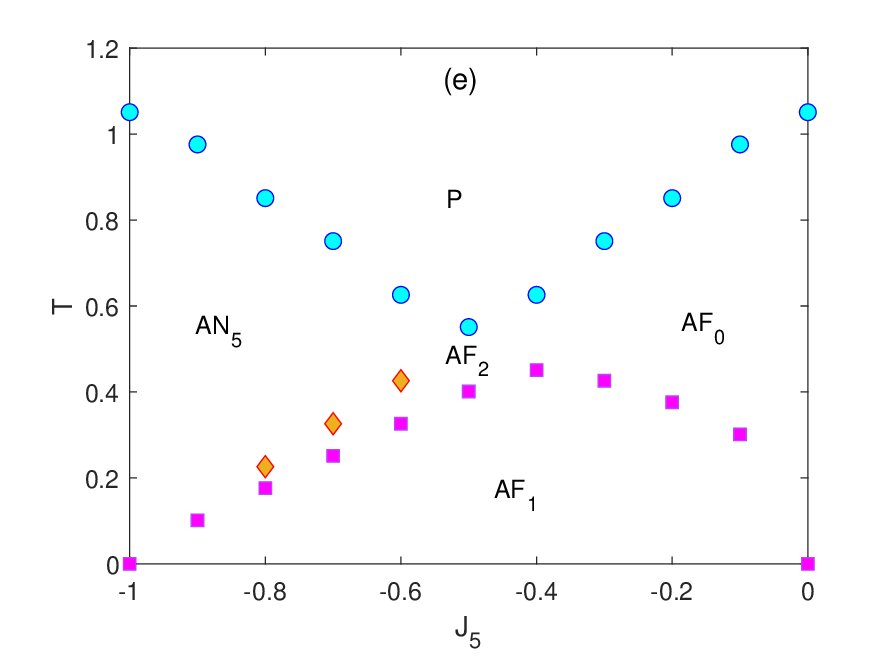}\label{fig:PD_q5}} \\
\subfigure{\includegraphics[scale=0.55,clip]{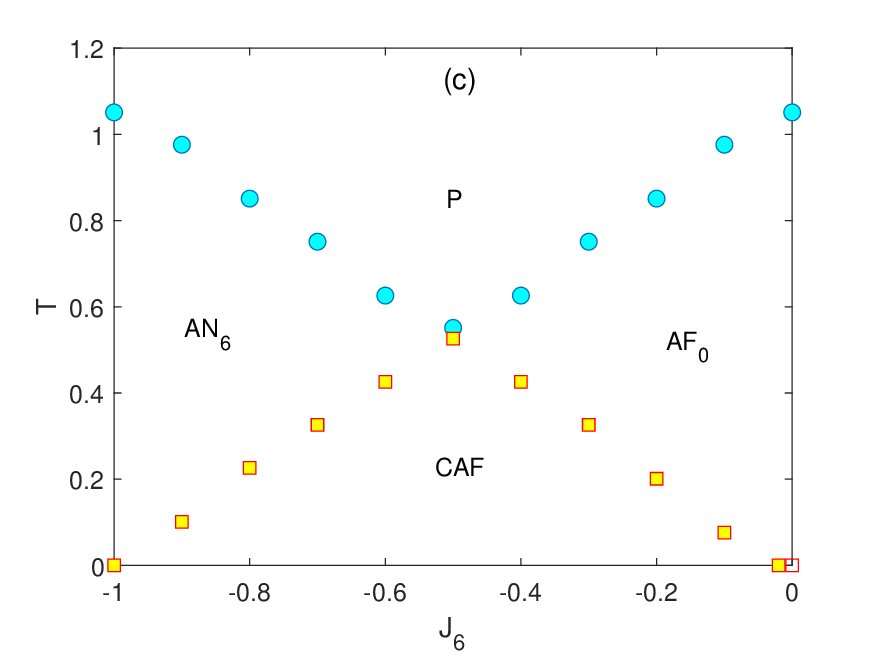}\label{fig:PD_q6}}
\subfigure{\includegraphics[scale=0.55,clip]{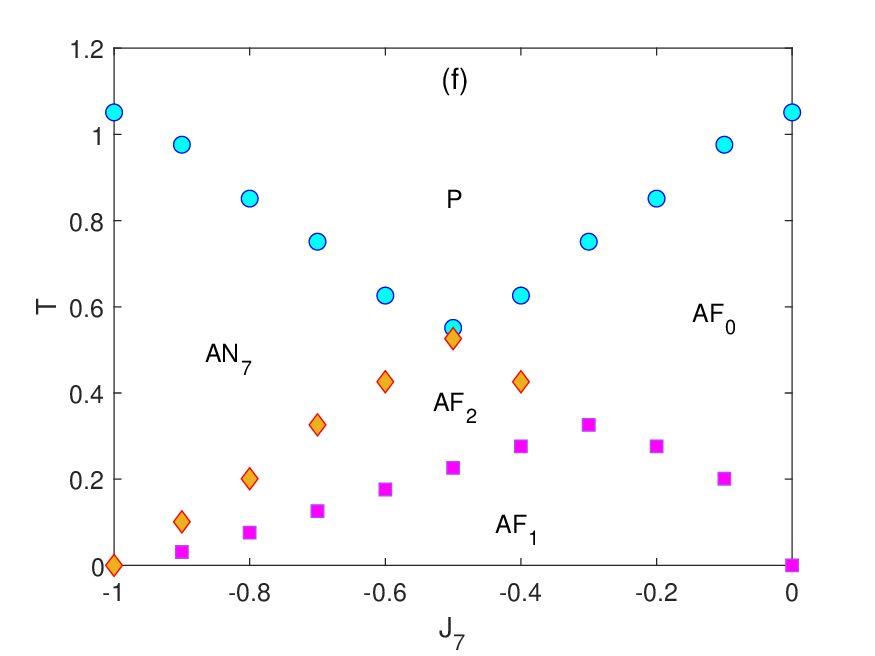}\label{fig:PD_q7}}
\caption{(Color online) Phase diagrams in $T-J_q$ planes, for $q=2,\hdots,7$. The insets in (b) show one-sublattice spin distributions at $T=0.15$ and $J_4=-0.9$ (AN$_4$), $J_4=-0.5$ (CAF), and $J_4=-0.1$ (AF$_0$).}\label{fig:PD}
\end{figure*}

The phase diagrams in the $T-J_q$ planes are presented in Fig.~\ref{fig:PD}, for $q = 2, 3, \hdots, 7$. One can notice three different types of topology, featuring from three up to five phases. The phase diagrams corresponding to even values of $q$ (left column) share the same topology with four phases: the high-temperature paramagnetic (P) phase, the AF$_0$ phase~\footnote{AF$_0$ corresponds to the standard AF phase, also observed in the absence of the AN$_q$ coupling. The index $0$ is only attached to unify the notation with the phases AF$_1$ and AF$_2$ that appear for odd values of $q \geq 5$.} for small $|J_q|$ values (dominant $J_1$ coupling), the AN$_q$ phase for large enough $|J_q|$, and the low-temperature CAF phase wedged between the AF$_0$ and AN$_q$ phases. 

On the other hand, the phase diagrams corresponding to odd values of $q$ (right column) display two types of topology. In particular, the $q=3$ case features three phases. Compared to the $q=2$ case, there is no CAF phase and the corresponding low-temperature region of the phase diagram is occupied by the AF$_0$ phase. It is also worth noticing that the absence of the competition between the two couplings for $q=3$ (that led to the formation of the CAF phase for $q=2$) also results in the overall shift of the phase boundaries to higher temperatures as well as the shift of the multicritical point, at which all the phase boundaries meet, to $J_q \approx -0.5$. The phase diagram topology changes for odd values of $q \geq 5$. More specifically, the AF$_0$ phase observed for $q=3$ splits into three different AF phases: AF$_0$, AF$_1$, and AF$_2$. The nature of these phases is very similar to the corresponding phases F$_0$, F$_1$, and F$_2$, reported in the F-N$_q$ models~\cite{pode11,cano14,cano16}, except for their AF character. In other words, in the present models, the F$_0$, F$_1$, and F$_2$ phases can be observed on each of the two AF-coupled sublattices. The respective phases can be on each sublattice characterized by multimodal spin angle distribution (AF$_2$) or unimodal distributions with smaller (AF$_1$) and larger (AF$_0$) widths. With the increasing $q$, the AF$_2$ phase area expands at the cost of the AF$_1$ phase. In the following, the respective phases and phase transitions are described in more detail for the representative cases.

\subsubsection{Even values of $q$}
The focus point of the case of even values of $q$ are the low-temperature phase transitions from the AF$_0$ and AN$_q$ phases to the CAF phase. Let us recall that a related CFM phase has been observed in the FM-AN$_2$ model, wedged between the FM and AN$_2$ phases~\cite{dian11,zuko19}. In the CFM phase, neighboring spins in two sublattices are canted by the interaction ratio dependent angle and the intrasublattice ordering is characterized by a fast-decaying power-law  correlation function. Significantly diminished correlations within CFM are argued to be a result of the presence of zero-energy domain walls due to the inherent degeneracy caused by the AN$_2$ interactions. The AN$_2$-CFM phase transition appears to be not compliant
with the Ising universality class due to the competition between the two types of couplings. However, the transition occurs at rather low temperature and is accompanied with huge MC autocorrelation times, which make a reliable FSS analysis problematic~\cite{zuko19}.

\begin{figure*}[t!]
\centering
\subfigure{\includegraphics[scale=0.55,clip]{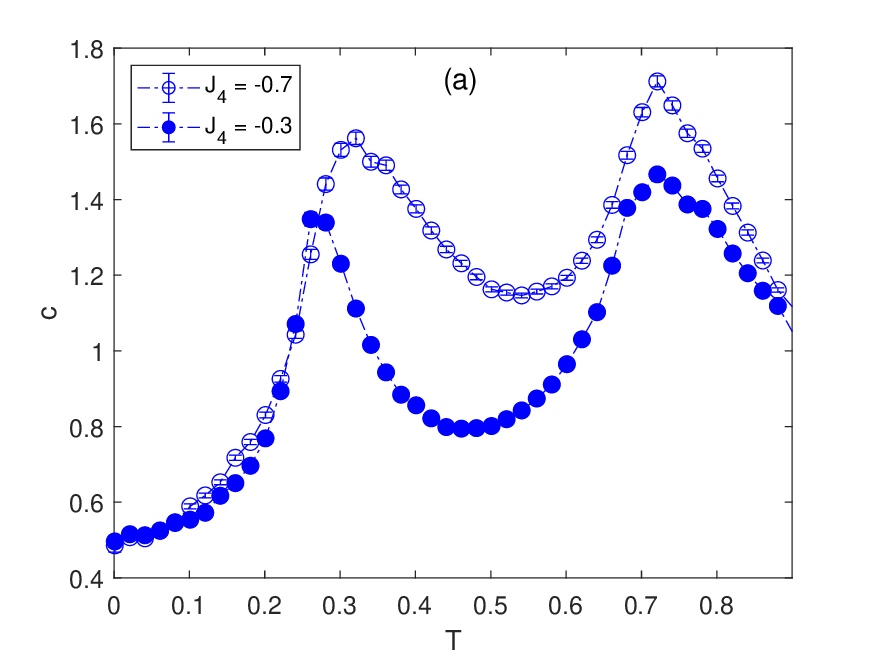}}\label{fig:T-c_J1_03_07_q4_err_an_L72}
\subfigure{\includegraphics[scale=0.55,clip]{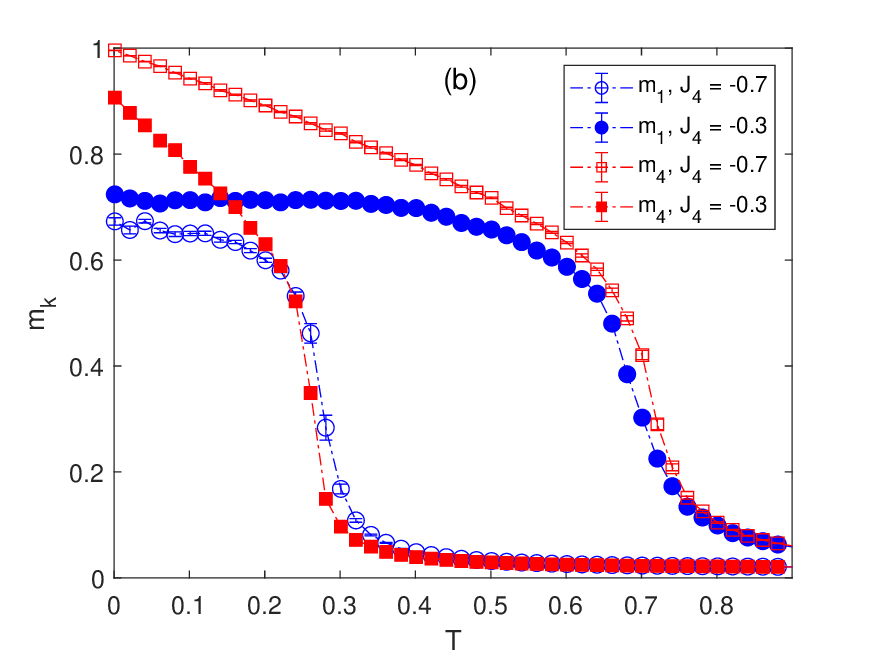}}\label{fig:T-mq_J1_03_07_q4_err_an_L72}\\
\subfigure{\includegraphics[scale=0.55,clip]{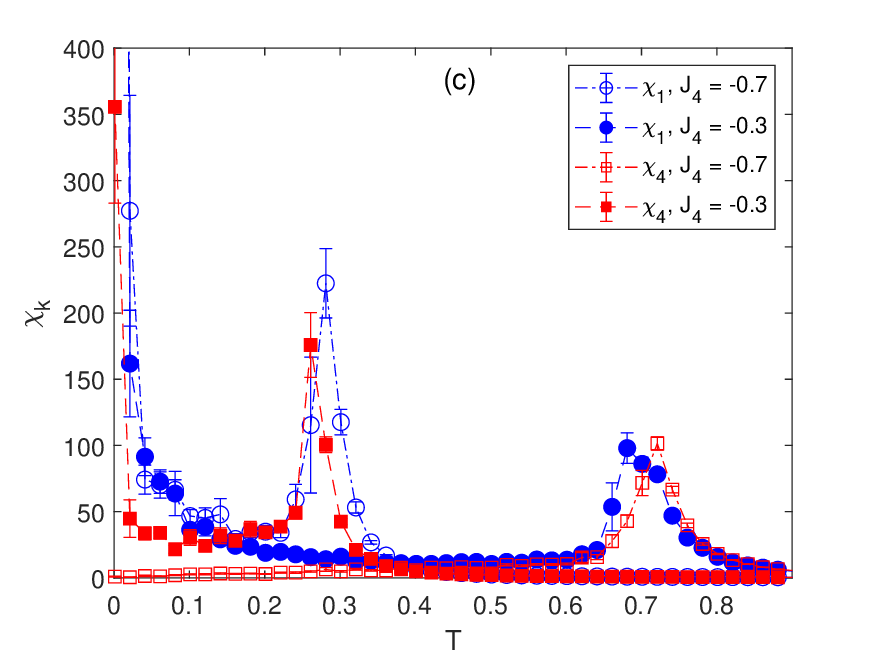}}\label{fig:T-chiq_J1_03_07_q4_err_an_L72}
\subfigure{\includegraphics[scale=0.55,clip]{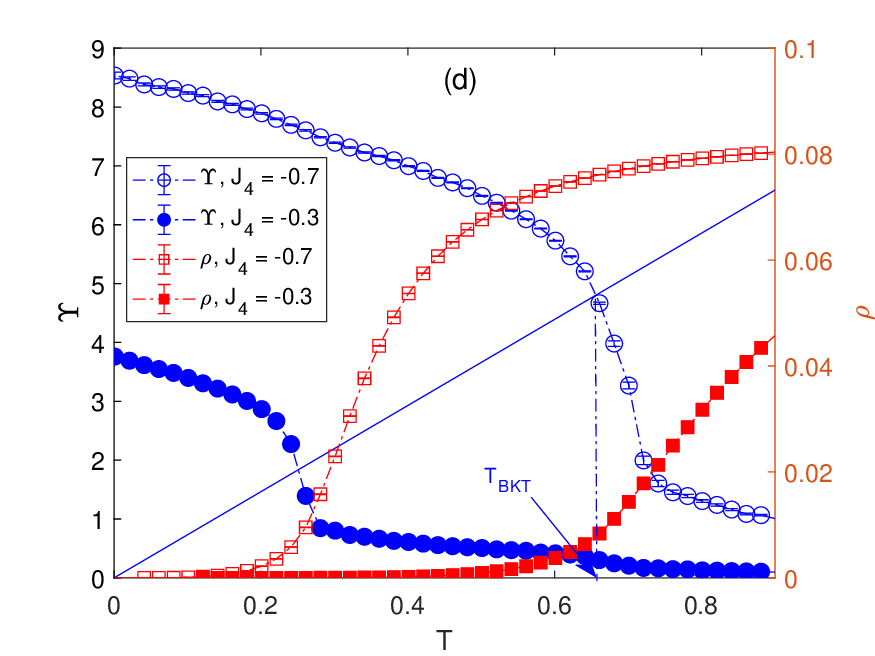}}\label{fig:T-hm_J1_03_07_q4_err_an_L72}
\caption{(Color online) Temperature dependencies of (a) the specific heat, (b) the magnetizations $m_1$ and $m_4$, (c) the susceptibilities $\chi_1$ and $\chi_4$ and (d) the helicity modulus $\Upsilon$ and vertex density $\rho$, for $q=4$ and $J_4=-0.3$ and $-0.7$. The straight line is $2(|J_1|+16|J_4|)T/\pi$, for $J_4=-0.7$.}\label{fig:q4}
\end{figure*}

For the present models with even values of $q$, the behavior of various evaluated thermodynamic quantities are presented in Fig.~\ref{fig:q4} for the representative case of $q=4$ (for which the transition temperatures to the CAF phase are much higher and the autocorrelation time much smaller than for the $q=2$ case). The temperature dependencies of those quantities are presented for two values of the parameter $J_4$, that correspond to the AF$_0$-CAF ($J_4=-0.3$) and the AN$_4$-CAF ($J_4=-0.7$) transitions. The specific heat curves in the panel (a) indicate the presence of two transitions in either case. The decay of the order parameters $m_1$ and $m_4$ in the panel (b) tell us that the high-temperature peaks correspond to the AF$_0$-P transition for $J_4=-0.3$ and the AN$_4$-P transition for $J_4=-0.7$. The low-temperature CAF-AF$_0$ transition for $J_4=-0.3$ and CAF-AN$_4$ transition for $J_4=-0.7$ are captured by the parameters $m_4$ and $m_1$, which remain finite in the CAF phase but vanish at the transitions to the AF$_0$ and AN$_4$ phases, respectively. The respective transition points can be roughly determined based on the peaks in the corresponding susceptibilities, $\chi_1$ and $\chi_4$ in the panel (c). One can notice that they occur at slightly lower temperatures than the corresponding specific heat peaks, as usually happens in the case of the BKT transition.

A true order parameter for the BKT transition is the helicity modulus $\Upsilon$. At the transition temperature $T_{\rm BKT}$ it exhibits a universal jump from a finite value to zero. In the standard $XY$ model $T_{\rm BKT}$ can be determined from the condition $\Upsilon(T_{\rm BKT}) = 2T_{\rm BKT}/\pi\nu^2$, where $\nu$ is the vorticity. In the generalized $XY$ model with the mixed vorticities, like the present one with $\nu_1=1$ and $\nu_2=1/q$, the simplest, albeit not very precise~\footnote{This assumption violates the universality of the stiffness
jump predicted within the BKT scenario~\cite{hubs13}.}, approach interpolates between the two pure cases obtaining $\Upsilon(T_{\rm BKT}) = (2/\pi) T_{\rm BKT}(|J_1|/\nu_1^{2}+|J_q|)/\nu_2^{2})$~\cite{qi13,park08,qin09}. The helicity modulus curves, along with the straight line showing the estimated BKT transition temperature for $J_4=-0.7$, are presented in Fig.~\ref{fig:q4}(d). The same figure also includes the (integer) vertex density $\rho$ curves for the two values of $J_4$. The vortex density is small within the BKT phase but as vortices unbind close to the transition to the disordered phase, their density rapidly increases. No increase is observed at the CAF-AF$_0$ transition, which means that the integer vortices remain bounded also inside the CAF phase.

\begin{figure*}[t!]
\centering
\subfigure{\includegraphics[scale=0.55]{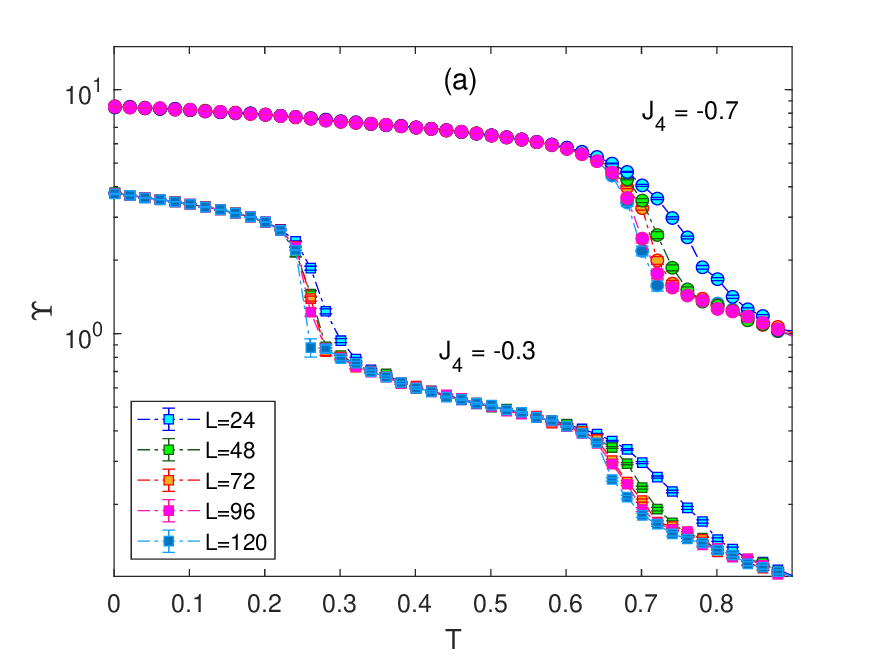}}\label{fig:T-hm_J1_0_3_07_q4_err_an_L24-120}
\subfigure{\includegraphics[scale=0.55]{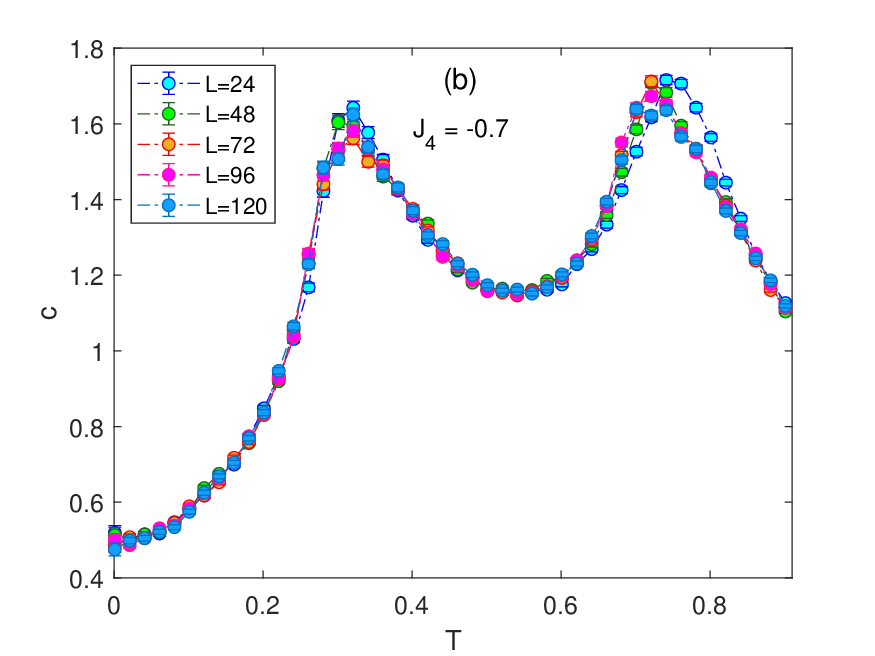}}\label{fig:T-c_J1_0_3_q4_err_an_L24-120}\\
\subfigure{\includegraphics[scale=0.55]{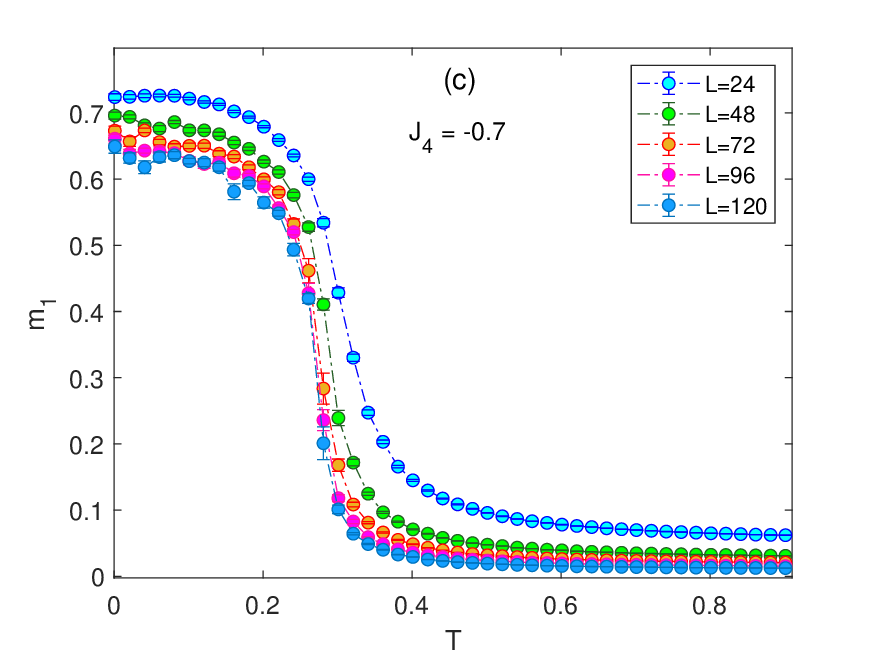}}\label{fig:T-m_J1_0_3_q4_err_an_L24-120}
\subfigure{\includegraphics[scale=0.55]{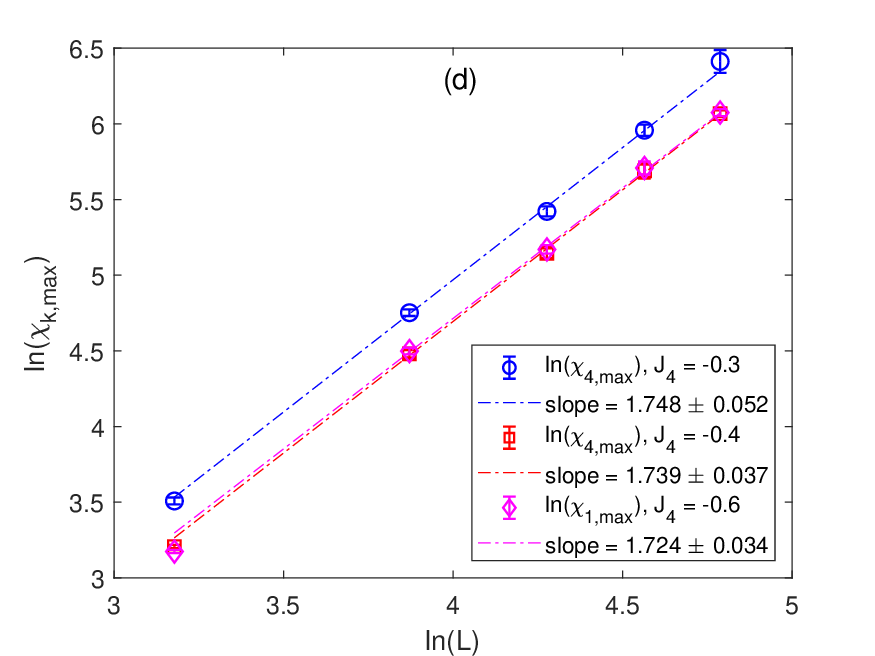}}\label{fig:fss_J1_04_06_07_q4}
\caption{(Color online) (a) Temperature dependencies of (a) the helicity modulus, (c) the specific heat, (c) the magnetization, for $q=4$, $J_4=-0.3$, $-0.7$ and different lattice sizes. (d) The FSS analysis of the susceptibility $\chi_1$ at the AF$_0$-CAF transition for $J_4=-0.6$ and the susceptibility $\chi_4$ at the AN$_4$-CAF transition for $J_4=-0.3$ and $-0.4$.}\label{fig:fss_q4}
\end{figure*}

Next we look into the nature of the respective phase transitions, particularly those to the peculiar CAF phase. To judge whether the transition has a BKT or non-BKT nature in Fig.~~\ref{fig:fss_q4} we plot the lattice size dependencies of various quantities to check their finite-size behaviors in the vicinity of the transition points. It is known that in the case of the BKT transition the helicity modulus should display a discontinuous jump in the thermodynamic limit and thus for finite lattice sizes one should observe a gradually increasing slope as the lattice size increases. In the panel (a) we plot the helicity modulus curves for different lattice sizes and two values of $J_4=-0.3$ and $-0.7$, corresponding to the CAF-AF$_0$ and CAF-AN$_4$ transitions, respectively. In the former case, in the vicinity of both transitions we indeed can see the size-dependent behavior, which indicates that these transitions are of the BKT nature. 

Nevertheless, for $J_4=-0.7$ the size-dependence is only observed at the high-temperature P-AN$_4$ transition. This may indicate a non-BKT nature of the low-temperature CAF-AN$_4$ transition. It is interesting to notice that this transition bears some similarity with the F$_1$-F$_2$ transition in the F-N$_q$ models for $q \geq 5$. At this transition spins in the phase F$_2$, equally distributed along $q/2$ directions in the same half-plane with peaks of unequal heights, pick in the F$_1$ phase one (the mostly populated) direction~\cite{cano16}. In the present model, as demonstrated in the insets in Fig.~\ref{fig:PD}(b), spins that belong to one sublattice are in the phase AN$_q$ equally distributed along $q$ directions around the whole plane and at the transition to the CAF phase choose one common direction (different from either of the $q$ directions in the AN$_q$ phase) that is canted with respect to the other sublattice spin direction by the $J_q$-dependent angle. Thus there is some resemblance with the $q=4$ clock model showing the Ising-like phase transition, except for the frustration-induced canting. This similarity along with the absence of the size-dependence of the helicity modulus, as defined in Eq.(~\ref{helicity}), at the CAF-AN$_4$ transition might suggest the non-BKT nature of the transition  but some other arguments would rather suggest the opposite. In particular, as has been shown by Kumano et al.~\cite{kuma13}, the helicity modulus insensitivity to the lattice size might be just the consequence of the fact that it is defined with respect to an infinitesimal twist, which is inappropriate not only for models with discrete $Z_p$ symmetry, such as the $p$-state clock model, but also for the $XY$ model with a perturbation which breaks the $U(1)$ symmetry down to $Z_p$. The lack of the jump in such improperly defined helicity modulus led to the false conclusion about the absence of the BKT transition in the five-state clock model~\cite{lapi06,baek13}. In favor of the BKT nature of the CAF-AN$_4$ transition are also the shape and the system size dependencies of the specific heat in the panel (b) and the magnetization in the panel (c). In particular, the low-temperature specific heat peak is not sharp, as it should be at the Ising transition, but rather round and similar to the high-temperature peak that accompanies the BKT transition. Furthermore, we verified the nature of the transitions by performing FSS analyses of maxima of the susceptibilities $\chi_1$ for $J_4=-0.6$, corresponding to the CAF-AN$_4$ transition, and $\chi_4$ for $J_4=-0.3$ and $-0.4$, corresponding to the CAF-AF$_0$ transition (see Fig.~\ref{fig:fss_q4}(d)). The fitted slopes~\footnote{To achieve the asymptotic power-law regime, the smallest size of $L=24$ was dropped out from the fits.} for the respective cases show only rather small variation and do not deviate from the value $2-\eta=7/4$ (expected at either Ising or BKT transition) beyond the statistical errors.

\begin{figure*}[t!]
\centering
\subfigure{\includegraphics[scale=0.55]{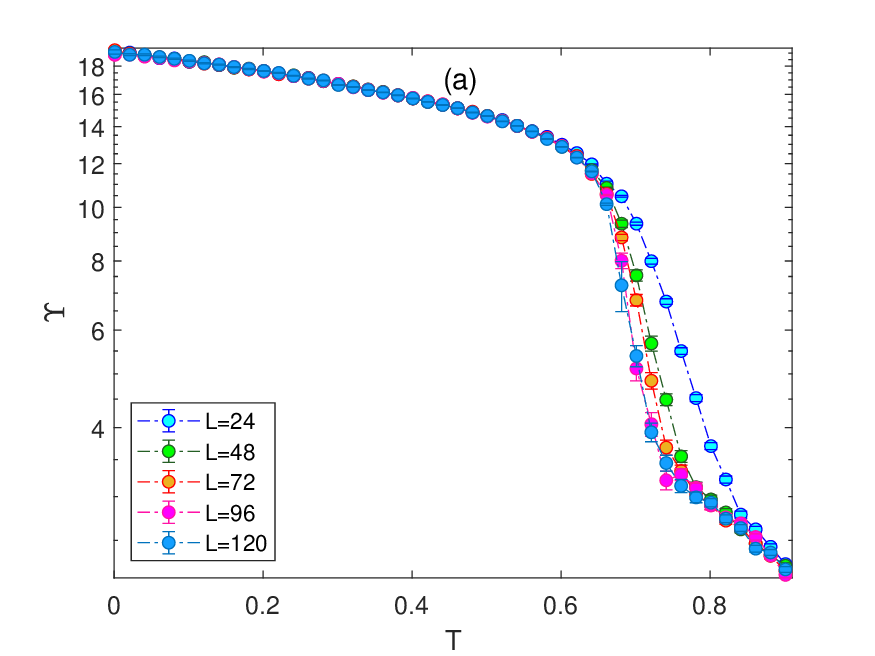}}\label{fig:T-hm_J1_0_3_07_q6_err_an_L24-120}
\subfigure{\includegraphics[scale=0.55]{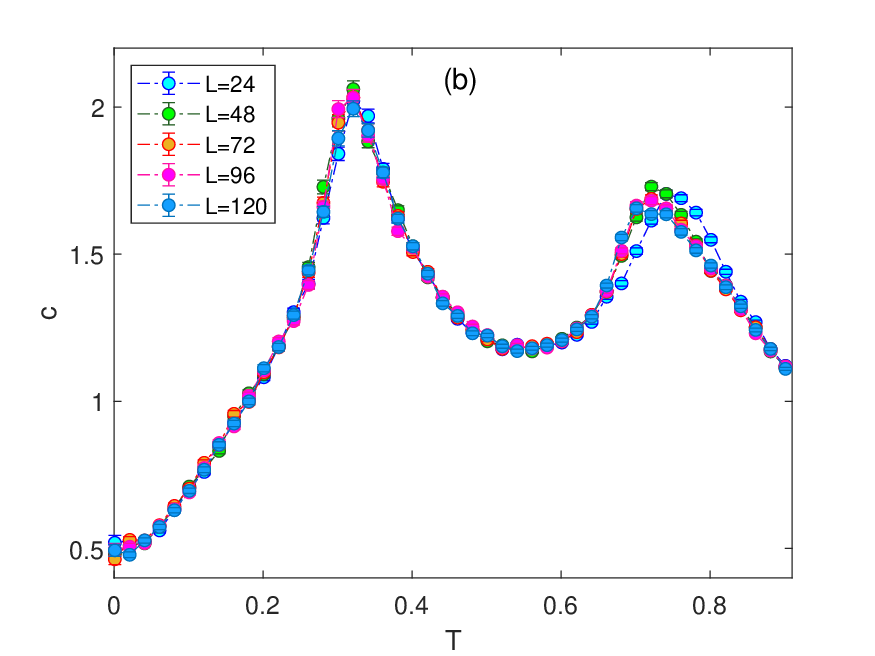}}\label{fig:T-c_J1_0_3_q6_err_an_L24-120}\\
\subfigure{\includegraphics[scale=0.55]{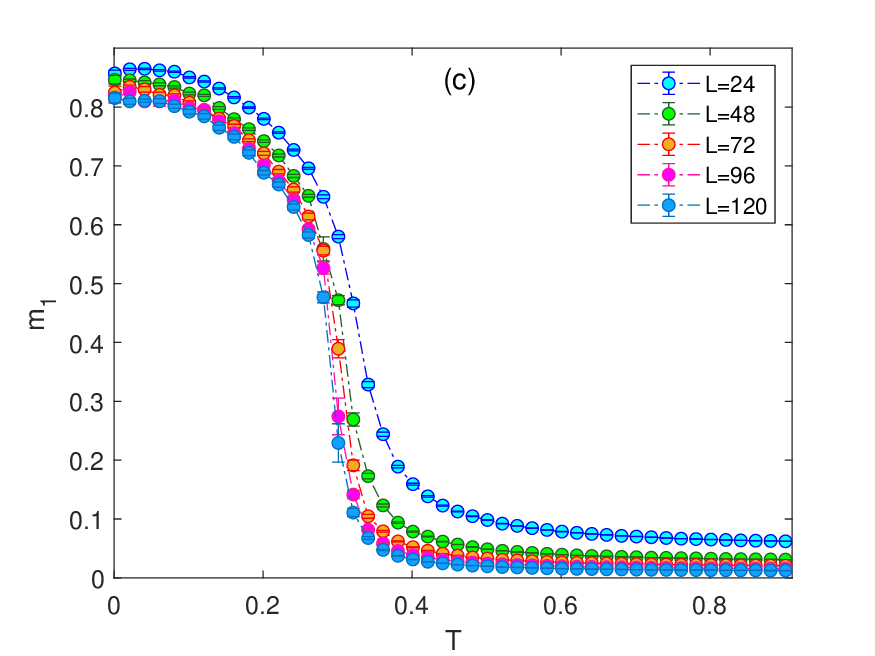}}\label{fig:T-m_J1_0_3_q6_err_an_L24-120}
\subfigure{\includegraphics[scale=0.55]{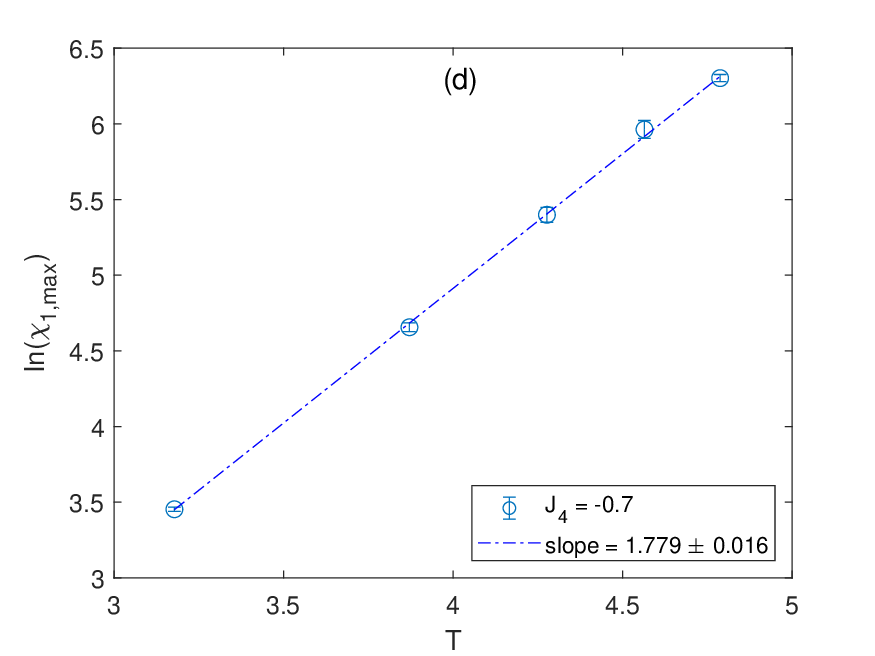}}\label{fig:fss_J1_03_q6_k1}
\caption{(Color online) (a) Temperature dependencies of (a) the helicity modulus, (b) the specific heat, (c) the magnetization, for $q=6$, $J_6=-0.7$ and different lattice sizes. (d) The FSS analysis of the susceptibility $\chi_1$ at the AF$_0$-CAF transition for $J_6=-0.7$.}\label{fig:fss_q6}
\end{figure*}

Similar examination of the disputable CAF-AN$_q$ transition was also performed for $q=6$. If the transition was related to the $q$-state clock model then we should observe the change from the Ising type for $q=4$ to the BKT nature for $q=6$. However, the results presented in Fig.~\ref{fig:fss_q6} appear consistent with those obtained for $q=4$ and thus we believe that it is more likely that both transitions are of the BKT nature.

\subsubsection{Odd values of $q$}

\begin{figure*}[t!]
\centering
\subfigure{\includegraphics[scale=0.55,clip]{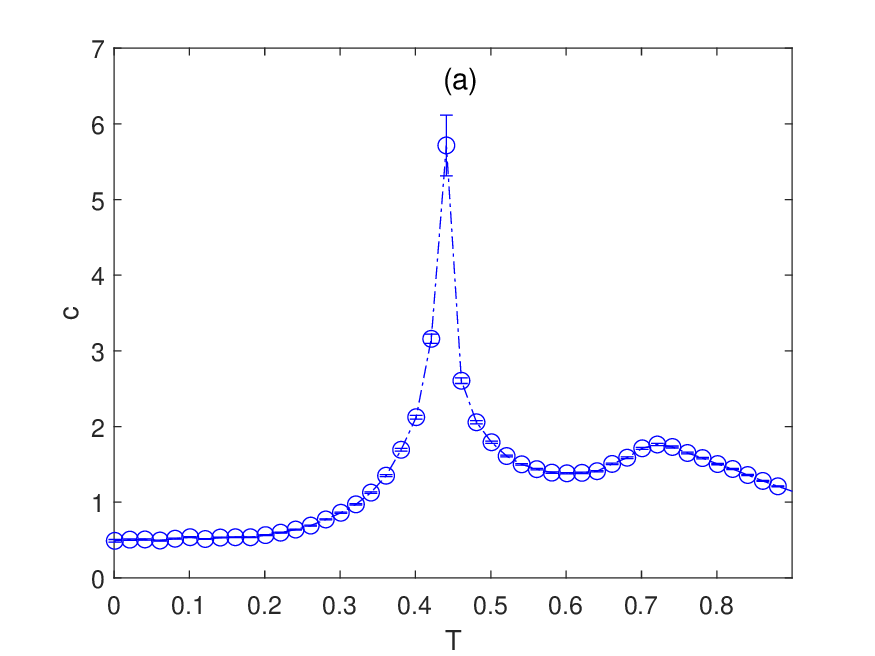}}\label{fig:T-c_J1_03_q3_L72}
\subfigure{\includegraphics[scale=0.55,clip]{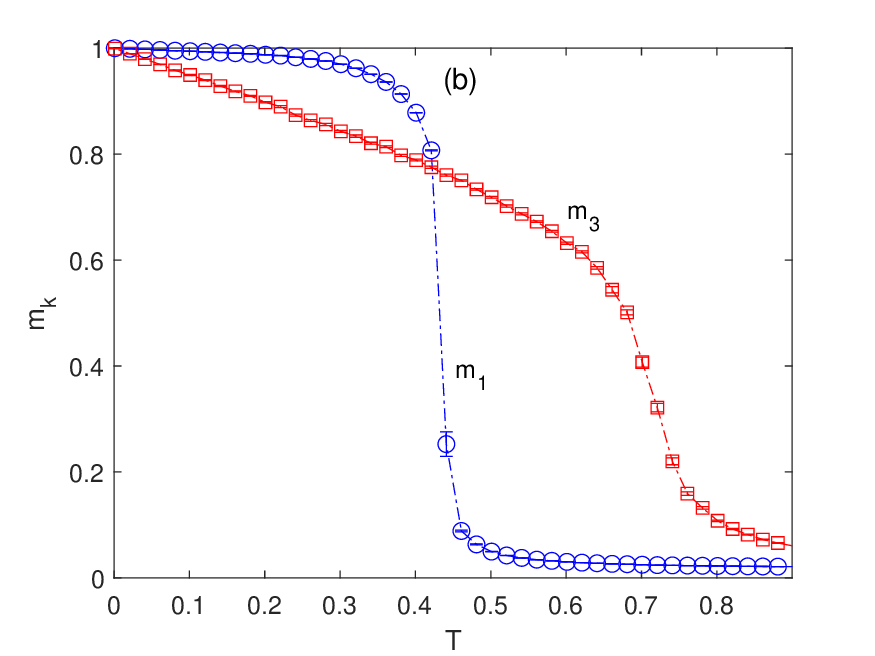}\label{fig:T-mk_J1_03_q3_L72}}\\
\subfigure{\includegraphics[scale=0.55,clip]{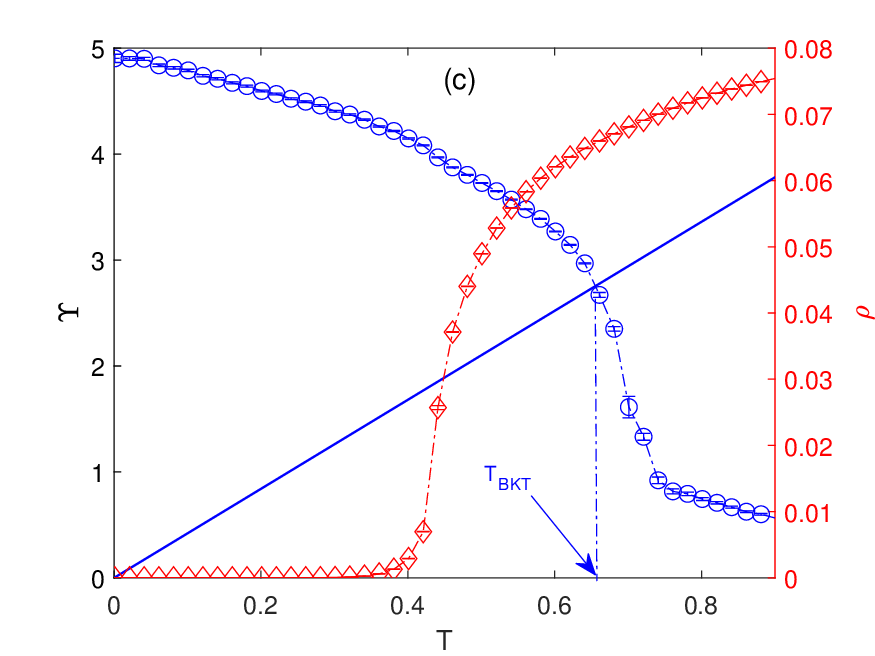}\label{fig:T-nv-hm_J1_03_q3_L72}}
\subfigure{\includegraphics[scale=0.55,clip]{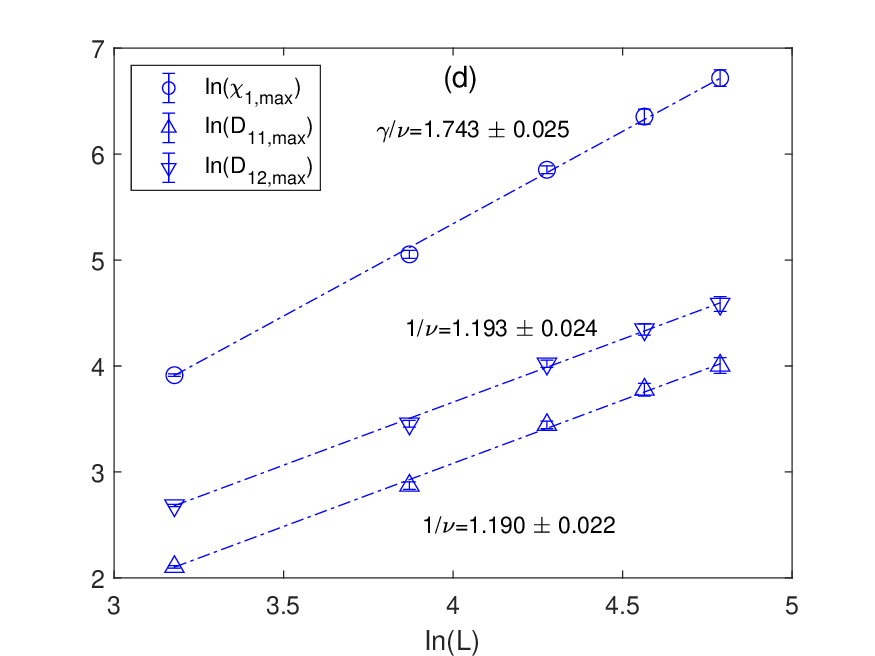}}\label{fig:fss_J1_03_q3}
\caption{(Color online) Temperature dependencies of (a) the specific heat, (b) the magnetizations $m_1$ and $m_3$, and (c) the helicity modulus $\Upsilon$ and the vortex density $\rho$ (the straight line is $2(|J_1|+q^2|J_3|)T/\pi$), for $q=3$ and $J_3=-0.7$. (d) The FSS analysis at the AN$_3$-AF$_0$ phase transition for $J_3=0.7$, with the estimated ratios of critical indices $\gamma/\nu$ and $1/\nu$.}\label{fig:q3}
\end{figure*}

For odd values of $q$, the case of $q=3$ is unique in the sense that its phase diagram features only two ordered phases: the AF$_0$ phase that occupies a vast area in the low-temperature region and the AN$_3$ phase that emerges above the AF$_0$ phase for $J_3 \lesssim -0.5$. The representative case from this region with $J_3=-0.7$ is presented in Fig.~\ref{fig:q3}. The temperature variation of the specific heat in the panel (a) signals the expected two phase transitions. While the high-temperature one is characterized by a round maximum, typical for the BKT transition, the low-temperature peak is sharp, suggesting a different kind of transition. The magnetizations $m_1$ and $m_3$, presented in the panel (b), confirm that the high-temperature transition is between the P and AN$_3$ phases and the low-temperature one is between the AN$_3$ and AF$_0$ phases. The vortex density and the helicity modulus temperature dependencies in the panel (c) demonstrate that the vortices unbind at the AN$_3$-AF$_0$ transition, which is accompanied with the sharp increase of the former and a small decrease of the latter quantity. Finally, the panel (d) presents the FSS analysis of the critical exponents at the low-temperature AN$_3$-AF$_0$ phase transition. The estimated values of the magnetic susceptibility and the correlation length exponents, $\gamma$ and $\nu$, are consistent with the three-states Potts universality class, as it was the case in the corresponding F-N$_3$ model~\cite{pode11,cano14}.

The phase diagram topology dramatically changes for odd values of $q>3$. Namely, the AF$_0$ phase, observed for $q=3$, splits into three phases AF$_0$, AF$_1$ and AF$_2$, as shown in Fig.~\ref{fig:PD}. Consequently, the system may show two or three phase transitions, depending on the value of $J_q$. The former scenario is realized for relatively small values of $|J_q|$. With the decreasing temperature the system first crosses from the P to the AF$_0$ phase and then to the AF$_1$ phase. At the increased values of $|J_q|$ but not exceeding the value of $|J_1|$, there is an intermediate phase AF$_2$, appearing between the AF$_0$ and AF$_1$ phases. In the regime of the dominant $|J_q|$, as the temperature decreases three phase transitions occur in the order P-AN$_q$, AN$_q$-AF$_2$ and AF$_2$-AF$_1$. 

\begin{figure*}[t!]
\centering
\subfigure{\includegraphics[scale=0.41,clip]{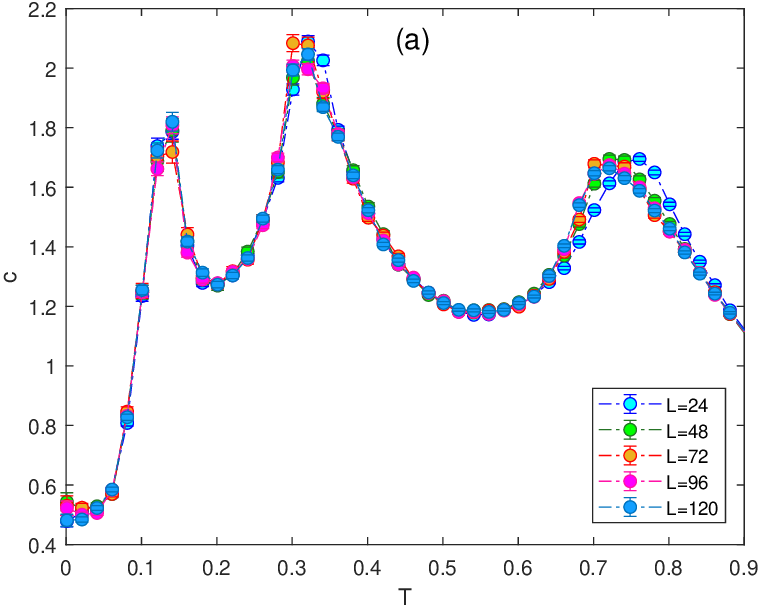}}\label{fig:T-c_J1_03_q7}
\subfigure{\includegraphics[scale=0.41,clip]{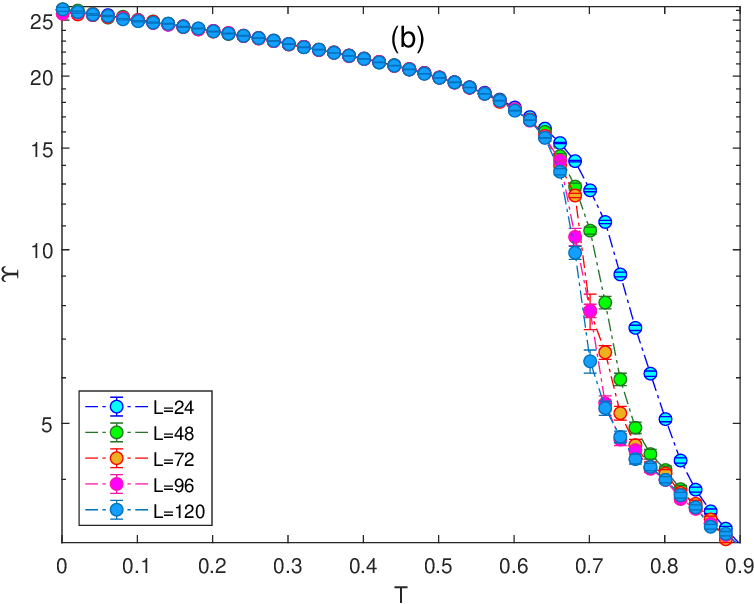}}\label{fig:T-hm_J1_03_q7}
\subfigure{\includegraphics[scale=0.41,clip]{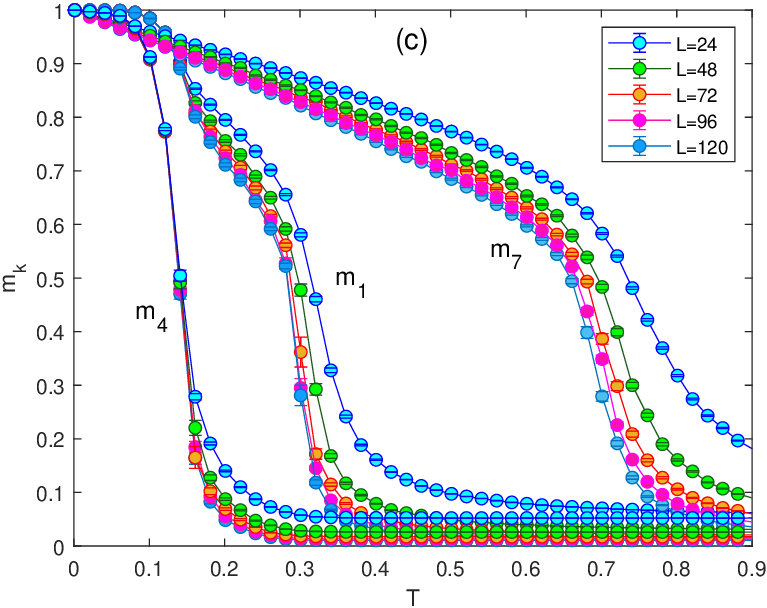}}\label{fig:T-m1_J1_4_7_03_q7}\\
\subfigure{\includegraphics[scale=0.41,clip]{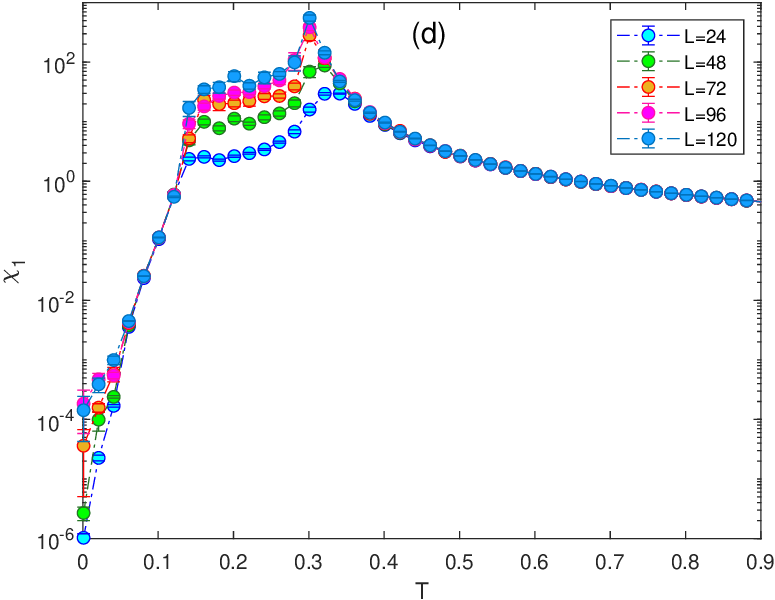}}\label{fig:T-chi1_J1_03_q7}
\subfigure{\includegraphics[scale=0.41,clip]{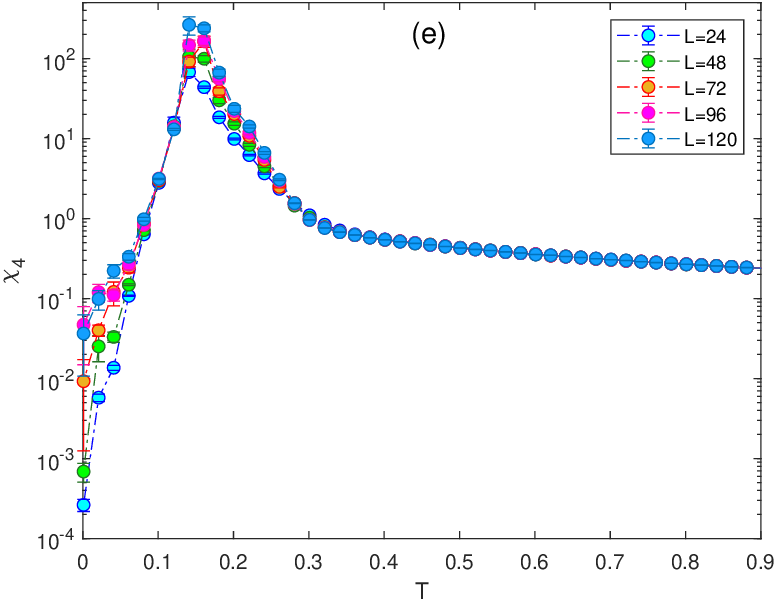}}\label{fig:T-chi4_J1_03_q7}
\subfigure{\includegraphics[scale=0.41,clip]{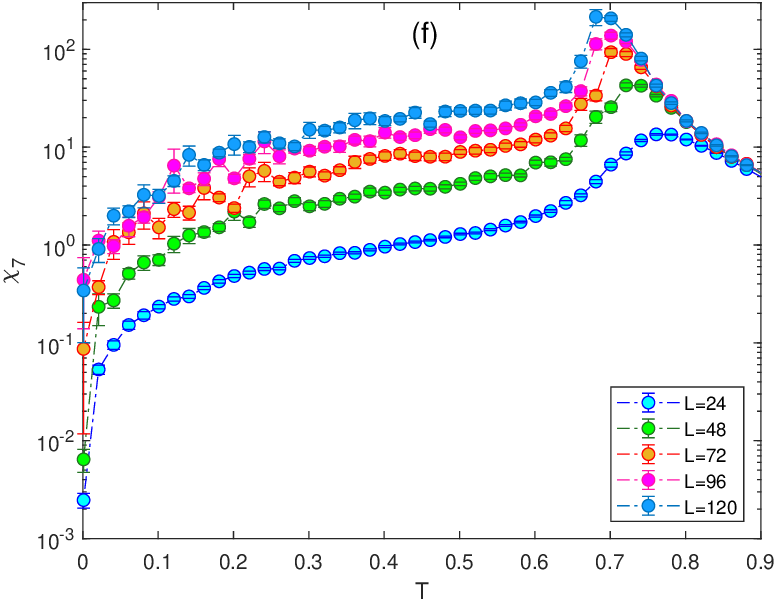}}\label{fig:T-chi7_J1_03_q7}
\caption{(Color online) Temperature dependencies of (a) the specific heat, (b) the helicity modulus, (c) the magnetizations $m_1$, $m_4$, and $m_7$, and (d-f) the susceptibilities $\chi_1$, $\chi_4$ and $\chi_7$, respectively, for $q=7$ and $J_7=-0.7$.}\label{fig:q7_J7_07}
\end{figure*}

\begin{figure*}[t!]
\centering
\subfigure{\includegraphics[scale=0.41,clip]{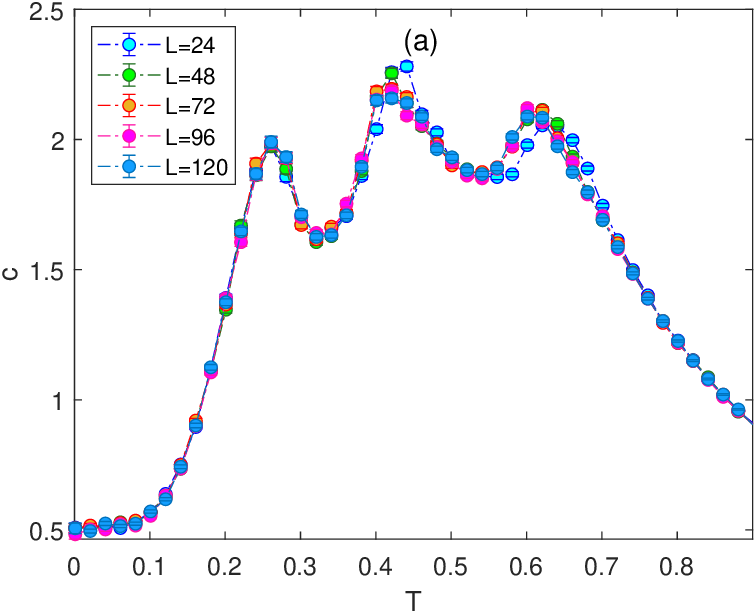}}\label{fig:T-c_J1_06_q7}
\subfigure{\includegraphics[scale=0.41,clip]{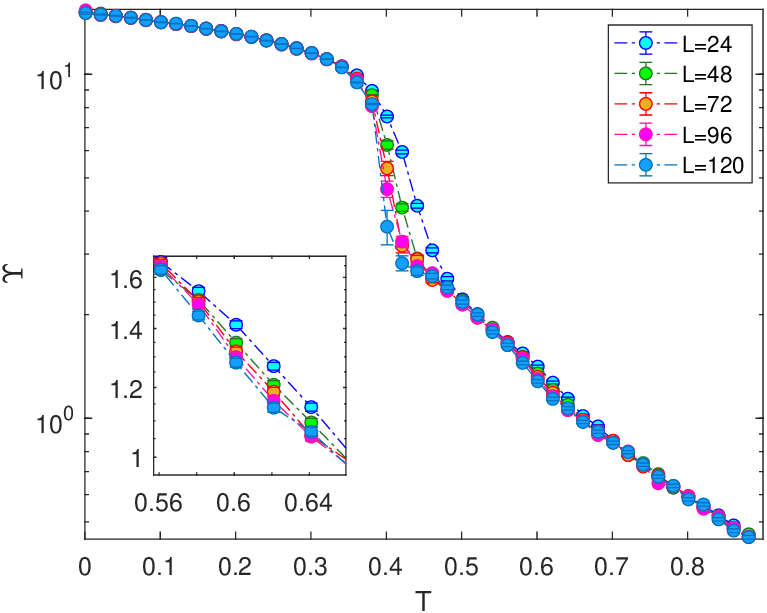}}\label{fig:T-hm_J1_06_q7}
\subfigure{\includegraphics[scale=0.41,clip]{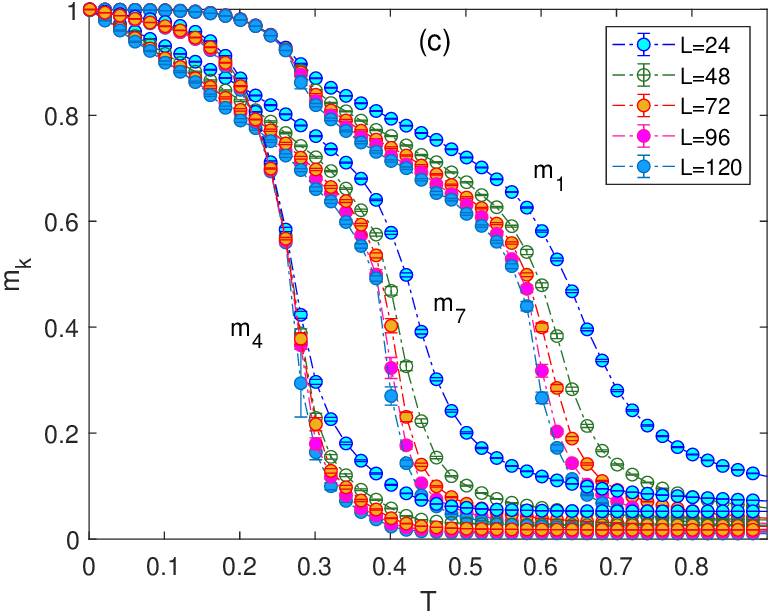}}\label{fig:T-m1_J1_4_7_06_q7}\\
\subfigure{\includegraphics[scale=0.41,clip]{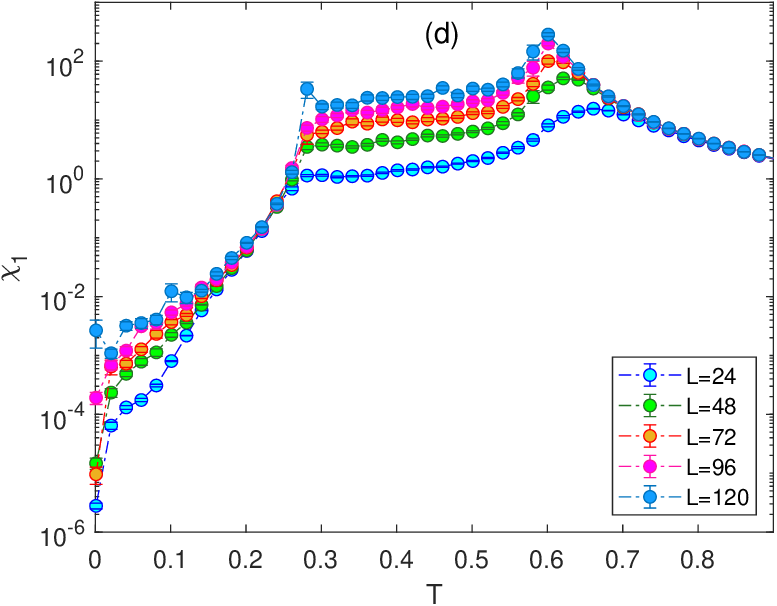}}\label{fig:T-chi1_J1_06_q7}
\subfigure{\includegraphics[scale=0.41,clip]{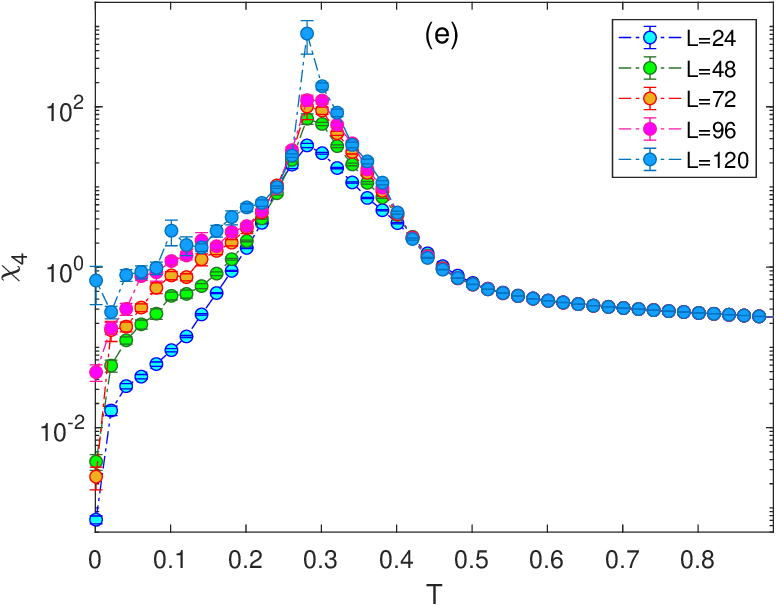}}\label{fig:T-chi4_J1_06_q7}
\subfigure{\includegraphics[scale=0.41,clip]{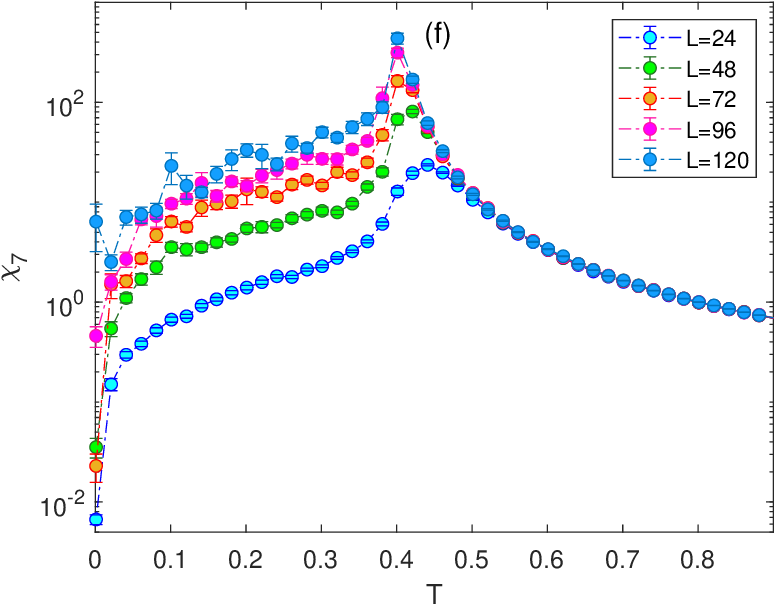}}\label{fig:T-chi7_J1_06_q7}
\caption{(Color online) Temperature dependencies of (a) the specific heat, (b) the helicity modulus, (c) the magnetizations $m_1$, $m_4$, and $m_7$, and (d-f) the susceptibilities $\chi_1$, $\chi_4$ and $\chi_4$, respectively, for $q=7$ and $J_7=-0.4$.}\label{fig:q7_J7_04}
\end{figure*}

\begin{figure*}[t!]
\centering
\subfigure{\includegraphics[scale=0.41,clip]{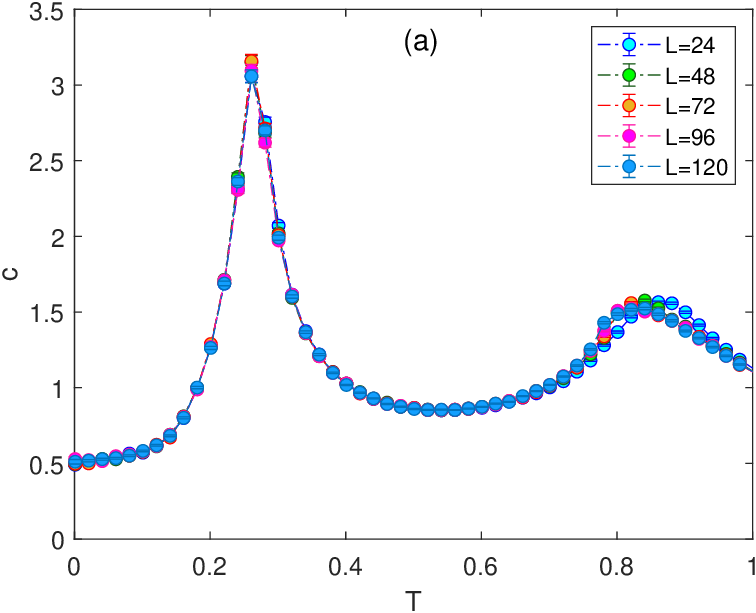}}\label{fig:T-c_J1_08_q7}
\subfigure{\includegraphics[scale=0.41,clip]{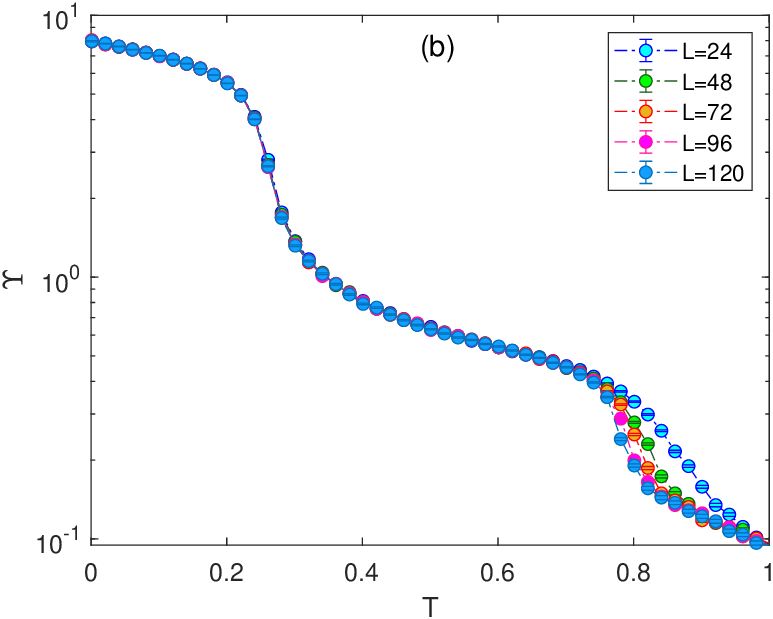}}\label{fig:T-hm_J1_08_q7}
\subfigure{\includegraphics[scale=0.41,clip]{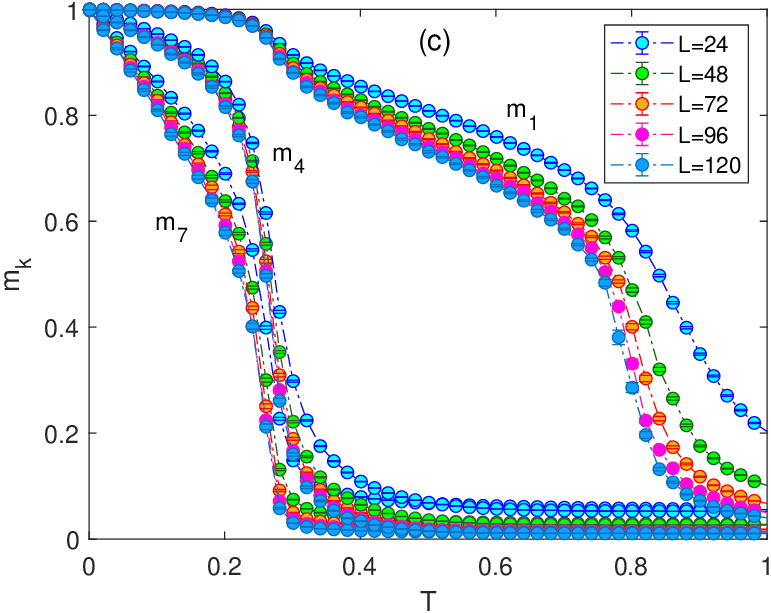}}\label{fig:T-mk_J1_08_q7}\\
\subfigure{\includegraphics[scale=0.41,clip]{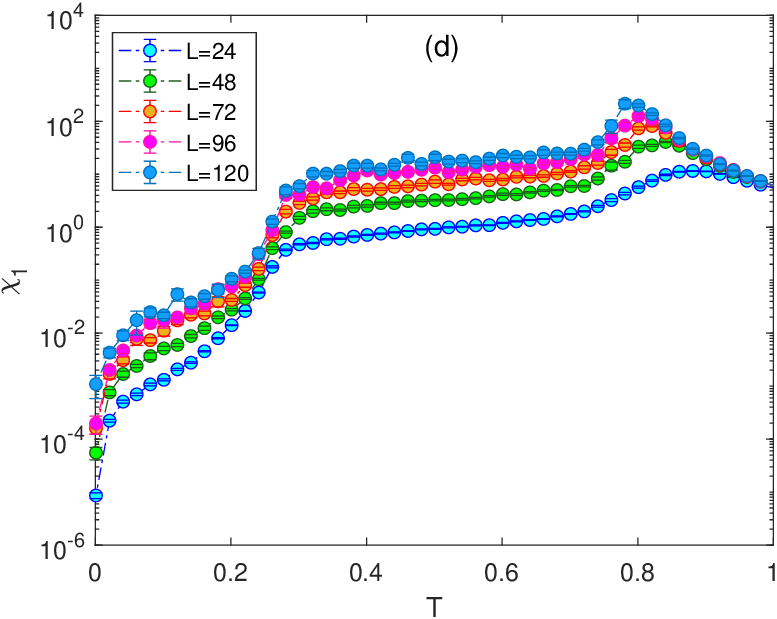}}\label{fig:T-chi1_J1_08_q7}
\subfigure{\includegraphics[scale=0.41,clip]{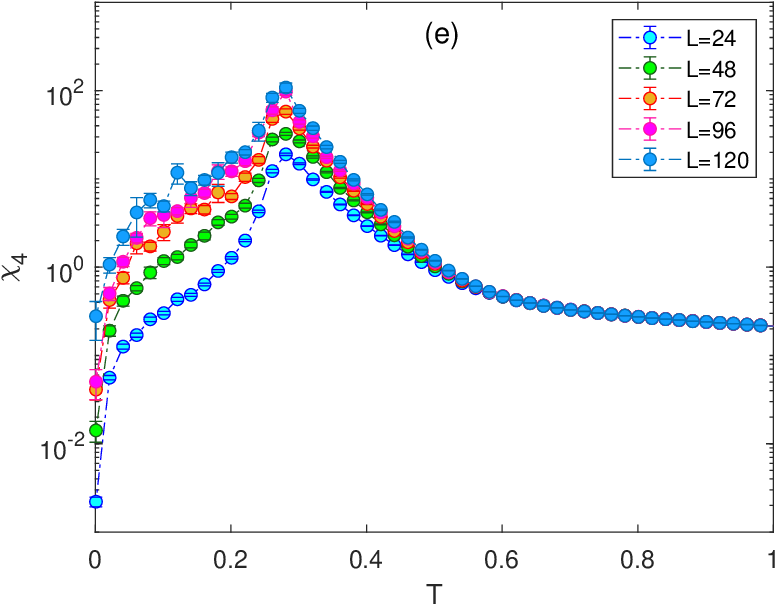}}\label{fig:T-chi4_J1_08_q7}
\subfigure{\includegraphics[scale=0.41,clip]{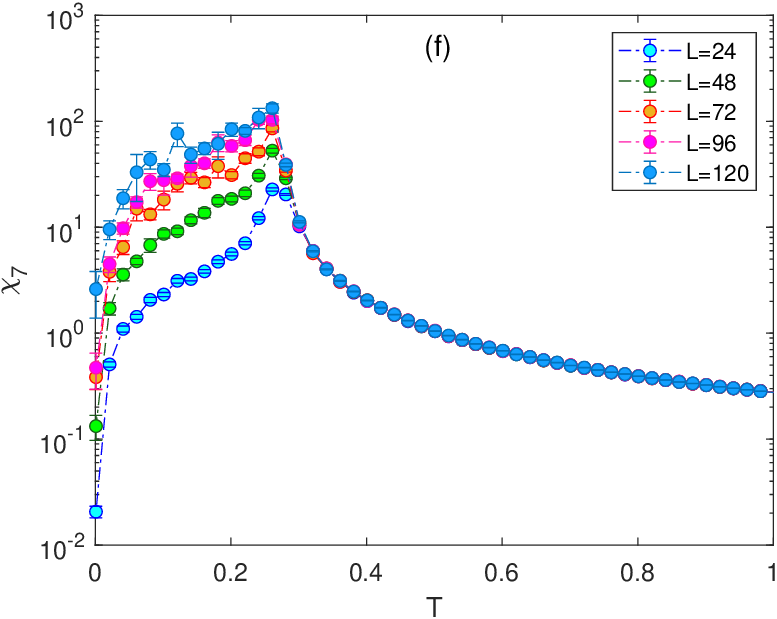}}\label{fig:T-chi7_J1_08_q7}
\caption{(Color online) Temperature dependencies of (a) the specific heat, (b) the helicity modulus, (c) the magnetizations $m_1$, $m_4$, and $m_7$, and (d-f) the susceptibilities $\chi_1$, $\chi_4$ and $\chi_4$, respectively, for $q=7$ and $J_7=-0.2$.}\label{fig:q7_J7_02}
\end{figure*}

The temperature variations of different quantities for all three scenarios are presented below, for $q=7$ with $J_7=-0.7$ (Fig.~\ref{fig:q7_J7_07}), $J_7=-0.4$ (Fig.~\ref{fig:q7_J7_04}) and $J_7=-0.2$ (Fig.~\ref{fig:q7_J7_02}). The plots include the curves corresponding to different system sizes, which can provide some insight into the nature of the respective phases and phase transitions. For the $J_7=-0.7$ case, the expected three phase transitions are demonstrated by the three peaks in the specific heat curves, presented in Fig.~\ref{fig:q7_J7_07}(a). The nature of the phases that they separate can be understood from the magnetization $m_k$ plots in the panel (c). In particular, the vanishing of $m_7$ signals the P-AN$_7$ transition, below which spins align along $q=7$ equally distributed directions in the plane with zero magnetization $m_1$. At lower temperature there is another AN$_7$-AF$_2$ transition. Within the AF$_2$ phase there is also a local alignment along several directions, but those are restricted to lie in the same half-plane leading to a finite magnetization $m_1$. Finally, the AF$_2$-AF$_1$ transition brings the system to the state with all spins narrowly aligned in the same direction and this transition can be detected by the decay of the magnetization $m_4$~\cite{cano16}.

Since the present models can show no true LRO~\cite{merm66}, all the observed phases have the QLRO BKT nature (they are critical at all temperatures within the phase). Consequently, the quantities $m_1$, $m_7$ and $m_4$ are not true order parameters of the respective phases and algebraically decay with the increasing system size, as demonstrated in Fig.~\ref{fig:q7_J7_07}(c). On the other hand, the corresponding susceptibilities show a power-law divergence governed by non-universal (temperature-dependent) critical exponents, as shown in the bottom panels of Fig.~\ref{fig:q7_J7_07}. Nevertheless, the transitions between these phases can also have a non-BKT nature, as for example evidenced in the F-N$_q$ models for the cases of $q=2$ and $3$ that show the second-order phase transitions belonging to the Ising and three-states Potts universality classes, respectively. One of the criteria for judging the nature of the respective transitions is the behavior of the helicity modulus. In the case of a BKT transition it is expected to show a jump at the transition point, which will be reflected by the size-dependent behavior in the vicinity of the transition temperature $T_{\rm BKT}$. On the other hand, no size dependence will be observed in the case of a non-BKT, such as the second-order, phase transition. The behavior of the helicity modulus presented in Fig.~\ref{fig:q7_J7_07}(b) suggests the BKT phase transition only for P-AN$_7$, while no size dependence can be observed at the remaining AN$_7$-AF$_2$ and AF$_2$-AF$_1$ transitions. This finding is consistent with the behavior of the F-N$_q$ model for $q > 4$, in which the N$_q$-F$_2$ transition was concluded to belong to the Ising universality, while the F$_2$-F$_1$ transition appeared to show a non-universal critical behavior or possibly just a crossover~\cite{cano16}.

Three phase transitions are also expected for the $J_7=-0.4$ case, as indicated in the specific heat curves in Fig.~\ref{fig:q7_J7_04}(a). However, the observed phases are different from the $J_7=-0.7$ case. In this case the local order parameters vanish in the following order: $m_1$ at the highest, $m_7$ at the intermediate and $m_4$ at the lowest temperature. This corresponds to the P-AF$_0$, AF$_0$-AF$_2$ and AF$_2$-AF$_1$ transitions, respectively. Like in Fig.~\ref{fig:q7_J7_07}, the size dependencies of the quantities $m_k$ and $\chi_k$, $k=1,4$ and $7$, within the respective phases indicate their BKT nature. However, in the helicity modulus curves in Fig.~\ref{fig:q7_J7_04}(b) we can observe two anomalies accompanied with the size-dependent behavior. Besides the P-AF$_0$ transition point (see a magnified view in the inset) a clear size dependence is also evident at the AF$_0$-AF$_2$ transition, which suggests that both P-AF$_0$ and AF$_0$-AF$_2$ transitions are of the BKT type. In contrast, no size dependence or even visible anomaly can be observed close to the AF$_2$-AF$_1$ transition. Thus, also in this area of the parameter space the picture of the phase transitions' nature in the present AF-AN$_q$ models is compatible with the one reported for the corresponding F-N$_q$ models~\cite{cano16}.

Finally, for $J_7 \gtrsim -0.3$ only two phase transitions can be observed, as demonstrated in Fig.~\ref{fig:q7_J7_02} for $J_7 = -0.2$. The high-temperature P-AF$_0$ transition is accompanied with the broad peaks in the specific heat curves and the lattice-size dependence of the helicity modulus, typical features observed at the BKT transition. On the other hand, the specific heat peaks at the low-temperature AF$_0$-AF$_1$ transition look much sharper and the helicity modulus shows no discernible dependence on the lattice size. These features point to a non-BKT nature of the AF$_0$-AF$_1$ transition, which is again in accordance with the findings in the F-N$_q$ models, which concluded this transition to be second order most likely in the Ising universality class~\cite{cano16}. It appears that for the AF$_0$-AF$_1$ transition either of the parameters $m_7$ or $m_4$ can be used (see Fig.~\ref{fig:q7_J7_04}(c)). However, the corresponding susceptibilities indicate that the appropriate one is $m_7$, since in $\chi_4$ the power-law dependence can be observed not only at the temperatures lower than the peak values but also within some range above the peak values. 

\begin{figure}[t!]
\centering
\includegraphics[scale=0.6]{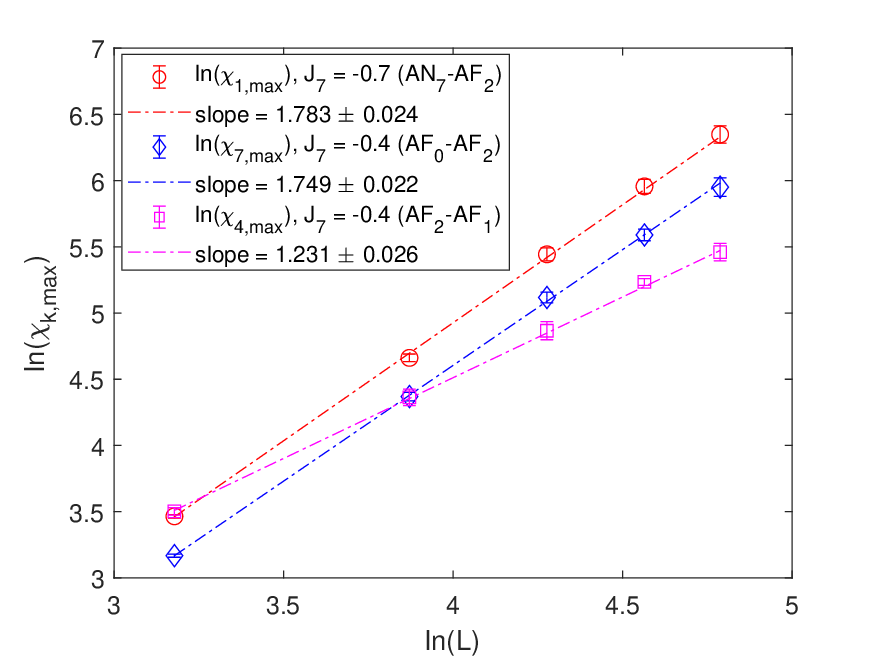}
\caption{(Color online) The FSS analysis of the susceptibility $\chi_1$ at the AN$_7$-AF$_2$ transition for $J_7=-0.7$, and the susceptibilities $\chi_7$ at the AF$_0$-AF$_2$ transition and $\chi_4$ at the AF$_2$-AF$_1$ transition, for $J_7=-0.4$.}\label{fig:fss_J1_03_AN7-AF2_06_AF0-AF2_q7}
\end{figure}

The observed phase transitions for odd values of $q>3$ are further inspected by performing FSS analyses for some representative cases. In Fig.~\ref{fig:fss_J1_03_AN7-AF2_06_AF0-AF2_q7} we present for $q=7$ the scaling of the appropriate susceptibilities $\chi_1$ associated with the AN$_7$-AF$_2$ transition for $J_7=-0.7$, and the susceptibilities $\chi_7$ at the AF$_0$-AF$_2$ and $\chi_4$ at the AF$_2$-AF$_1$ transitions, for $J_7=-0.4$. The slopes of the former two transitions display very similar values, compatible with $2-\eta = 7/4$, expected for either the Ising or BKT universality classes. As discussed above, the AN$_7$-AF$_2$ transition has non-BKT nature and is associated with a reflection symmetry breaking in the Ising universality class. At this transition the generally $q$ preferential directions evenly distributed around the circle on each sublattice in the AN$_q$ phase, reduce to one half distributed in the same half-plane in the AF$_2$ phase (see also discussion in Ref.~\cite{cano16} for $q=8$). On the other hand, in the AF$_0$-AF$_2$ transition the value of $\eta=1/4$ signifies the BKT type of the transition. The last case of the AF$_2$-AF$_1$ transition is again of non-BKT type and the estimated (non-universal) value of $2-\eta \approx 1.23$ is consistent with the results presented for the F$_2$-F$_1$ transition in the corresponding generalized ferromagnetic model~\cite{cano16}. Besides the above indications of the non-BKT nature, we also tried to perform the FSS analysis at the AF$_0$-AF$_1$ transition to identify the corresponding universality class. Nevertheless, as one can also observe in Fig.~\ref{fig:q7_J7_02}, fluctuations in the order parameter susceptibility $\chi_7$ are too large to obtain a reliable estimate of the critical exponent.

\section{Discussion and conclusion}

Critical properties of the generalized $XY$ model with the antiferromagnetic (AF) $J_1 \in (-1,0)$ and antinematic-like (AN$_q$) $J_q = -J_1-1 \in (-1,0)$ nearest-neighbor interactions on a square lattice were studied for $q > 2$. The purpose of this study was to extend and generalize our earlier work, that focused on the related FM-AN$_2$ model~\cite{zuko19}, and to confront our findings with those reported for the corresponding FM-N$_q$ models with the ferromagnetic (FM) $J_1 \in (0,1)$ and nematic-like (N$_q$) $J_q = 1-J_1 \in (0,1)$ terms~\cite{pode11,cano14,cano16}. 

Let us recall that the FM-N$_q$ model for $q = 2$ produced the phase diagram with the FM ground state (GS) for any ratio between $J_1$ and $J_2$ and separate FM and N$_2$ phases at finite temperatures. There is the usual BKT transition to the P phase but the FM-N$_2$ transition was confirmed to belong to the Ising universality class~\cite{lee85,kors85}. On the other hand, the critical behavior of the corresponding model with the AN$_2$ interaction was found to be much different~\cite{dian11,zuko19}. In particular, GS was FM only for sufficiently weak $J_2 \in (-0.2,0)$, while for $J_2 \in (-1,-0.2)$ the competition between the collinear FM and noncollinear AN$_2$ ordering tendencies resulted in an intricate highly-degenerate canted FM (CFM) GS. Consequently, for $J_2 \in (-0.2,0)$ only one P-FM phase transition occurs at finite temperatures. On the other hand, for $J_2 \in (-1,-0.2)$ there are two successive phase transitions: the first one is from P to either AN$_2$ or FM (depending on $J_2$) phase, followed by another one to the CFM phase at very low temperatures. 

This discrepancy between the $q=2$ models with N$_2$ and AN$_2$ couplings already indicated that the evolution of the phase diagram topologies of the generalized FM-N$_q$ and AF-AN$_q$ models with the parameter $q$ will not follow the same trajectory. Indeed, in the FM-N$_q$ models the topology of the phase diagram remained unchanged for $q = 3$ and $4$, with the FM-N$_3$ and FM-N$_4$ transitions belonging to the three-state Potts and Ising universality classes, respectively~\cite{pode11,cano14}. However, for $q \geq 5$, the topology of the phase diagram changed dramatically with two new competition-driven FM phases emerging at low temperatures. Some of the phase transitions showed the BKT signature, while others indicated a non-BKT nature~\cite{cano16}.

The present study showed that the phase diagrams of the two models are similar only for odd values of the parameter $q$ and the respective phases reported for the FM-N$_q$ models can be for a given $q$ observed in the AF-AN$_q$ models on each of the two AF-coupled sublattices. Our FSS analysis indicated that also the nature of the corresponding phase transitions is the same as in the FM-N$_q$ models. On the other hand, for even values of $q$ the phase diagrams of the AF-AN$_q$ models are different from the FM-N$_q$ models and their topology does not change with $q$. In particular, besides the pure AF$_0$ and AN$_q$ phases, observed at higher temperatures in the regions of the dominant respective couplings, at low temperatures there is a new CAF phase wedged between the AF$_0$ and AN$_q$ phases. The CAF phase results from the competition between the AF$_0$ and AN$_q$ ordering tendencies and has no counterpart in the FM-N$_q$ model. Both AF$_0$-CAF and AN$_q$-CAF phase transitions appear to have the BKT nature. Nevertheless, particularly the AN$_q$-CAF transition for even values of $q$ deserves further examination to provide more conclusive evidence.

The observed qualitative differences in the critical behavior of the AF-AN$_q$ models for even and odd values of the parameter $q$ suggest possible directions of the future research. For example, further generalization of the model to the Hamiltonian ${\mathcal H}=-\sum_{k=1}^{q}J_k\sum_{\langle i,j \rangle}\cos(k\phi_{i,j})$, which includes a mixture of both even and odd terms that compete with each other, may lead to novel critical behavior. In the related models that include up to infinite number of higher-order terms with alternating signs and exponentially vanishing strength it was recently found that their mutual competition and collaboration leads to the critical behavior that strongly depends on whether the number of terms is odd or even~\cite{zuko22}.

\begin{acknowledgments}
This work was supported by the grants of the Scientific Grant Agency of Ministry of Education of Slovak Republic (No. 1/0695/23) and the Slovak Research and Development Agency (No. APVV-20-0150).
\end{acknowledgments}

\end{document}